\documentclass[aps,amsmath,twocolumn,amssymb,floatfix,showpacs,superscriptaddress,nofootinbib,longbibliography]{revtex4-1}
\usepackage{mathtools}
\usepackage{braket}
\usepackage[dvipsnames]{xcolor}
\usepackage{float}
\usepackage{subfigure}
\usepackage{dsfont}
\usepackage{makecell}
\usepackage[version=4]{mhchem}

\usepackage{multirow}

\usepackage[colorlinks=true,linktoc=page,linkcolor=magenta,citecolor=blue,urlcolor=violet]{hyperref}

\mathchardef\mhyphen="2D 

\newcommand{\ie}{{i.e.,\,\,}}

\usepackage{xcolor}

\newcommand\bea{\begin{eqnarray}}
	\newcommand\eea{\end{eqnarray}}
\newcommand\beq{\begin{equation}}  
	\newcommand\eeq{\end{equation}}

\newcommand{\hb}{\hbar} 
\newcommand{\del}{\partial} 
\newcommand{\ttilde}{\tilde{t}} 
\newcommand{\Gzero}{G^{0}} 

\newcommand{\tauzero}{\tau_0}
\newcommand{\tax}{\tau_x}

\newcommand{\tauz}{\tau_z}

\newcommand{\szero}{\sigma_0}

\usepackage[normalem]{ulem}
\definecolor{lime}{HTML}{A6CE39}
\usepackage{sidecap,tikz}
\DeclareRobustCommand{\orcidicon}{\hspace{-1.0mm}
	\begin{tikzpicture}
		\draw[lime, fill=lime] (0.0,0.0) 
		circle [radius=0.15] 
		node[white] {{\fontfamily{qag}\selectfont \tiny \,ID}};
		\draw[white, fill=white] (-0.0525,0.095) 
		circle [radius=0.007];
	\end{tikzpicture}
	\hspace{-3.0mm}
}
\foreach \x in {A, ..., Z}{\expandafter\xdef\csname orcid\x\endcsname{\noexpand\href{https://orcid.org/\csname orcidauthor\x\endcsname}
		{\noexpand\orcidicon}}
}

\AtBeginDocument{%
	\newwrite\bibnotes
	\def\bibnotesext{Notes.bib}
	\immediate\openout\bibnotes=\jobname\bibnotesext
	\immediate\write\bibnotes{@CONTROL{REVTEX41Control}}
	\immediate\write\bibnotes{@CONTROL{%
			apsrev41Control,author="08",editor="1",pages="1",title="1",year="1"}}
	\if@filesw
	\immediate\write\@auxout{\string\citation{apsrev41Control}}%
	\fi
}%
\begin{document}


	\title{Proximity-induced superconductivity and emerging topological phases in altermagnet-based heterostructures} 
	
	\author{Ohidul Alam \orcidA{}}
	\affiliation{Institute of Physics, Sachivalaya Marg, Bhubaneswar-751005, India}
	\affiliation{Homi Bhabha National Institute, Training School Complex, Anushakti Nagar, Mumbai 400094, India}
	
	\author{Amartya Pal \orcidB{}}
	\affiliation{Institute of Physics, Sachivalaya Marg, Bhubaneswar-751005, India}
	\affiliation{Homi Bhabha National Institute, Training School Complex, Anushakti Nagar, Mumbai 400094, India}
	
	\author{Paramita Dutta \orcidD{}}\thanks{PD and AS jointly conceived the idea and supervised the project.}
	\affiliation{Theoretical Physics Division, Physical Research Laboratory, Navrangpura, Ahmedabad - 380009, India}
	
	\author{Arijit Saha \orcidC{}}\thanks{PD and AS jointly conceived the idea and supervised the project.}
	\affiliation{Institute of Physics, Sachivalaya Marg, Bhubaneswar-751005, India}
	\affiliation{Homi Bhabha National Institute, Training School Complex, Anushakti Nagar, Mumbai 400094, India}
	

	\begin{abstract}
		We present a theoretical framework for investigating superconducting proximity effect in altermagnet (AM)–superconductor (SC) heterostructures. In general, AMs, characterized by vanishing net magnetization but spin-split electronic spectra, provide a promising platform for realizing unconventional quantum phases. We consider a two-dimensional $d$-wave AM proximity coupled to a three dimensional ordinary $s$-wave SC. By integrating out the superconducting degrees of freedom, we derive an effective Hamiltonian that describes the proximity-induced modifications in the AM layer in the form of a self-energy. We then derive an effective Green’s function to obtain the proximity-induced pairing amplitudes in the AM layer and classify the induced pairing amplitudes according to their parity, frequency, and spin. We find the presence of even-parity singlet and triplet pairing amplitudes in the AM layer. To achieve the odd-parity triplet components, important to realize topological superconductivity, we introduce a layer of Rashba spin-orbit coupling (RSOC) in the heterostructure. We analyse the band topology of this proximity-induced AM-RSOC layer and demonstrate the emergence of both weak and strong topological superconducting phases with edge-localized modes, characterized by winding number and Chern number. These findings highlight the role of AM–SC hybrid setup as a versatile platform for realizing odd-parity triplet pairings and engineering topological superconductivity in two dimensions.
	\end{abstract}

	\maketitle

	\section{Introduction}
	
	The recent discovery of a novel class of magnetic materials, termed as altermagnets (AMs), lead to a plethora of research activity over the past few years ~\cite{Smejkal_PRX_1,Smejkal_PRX_2,Mazin_PRX_2022,BhowalPRX2024,Mazin_PRBL_2023}. These AMs feature collinear compensated magnetic order where the opposite spin sublattices are related via rotation rather than translation or inversion. Unlike the conventional antiferromagnets, the broken time-reversal symmetry (TRS) in AMs leads to momentum dependent spin-splitting of the electronic bands inspite of vanishing net magnetization. In two dimensions (2D), the most commonly used AM is the $d$-wave AM where the magnetic order mimics the shape of planar $d$-wave orbitals with the up and down spin sectors are connected via $C_4$ rotation about the axis perpendicular to the plane. Candidate materials for such AMs include RuO$_2$~\cite{Smejkal_PRX_1,Smejkal_PRX_2,XZhou2024_PRL,Mazin_PRX_2022,Bai_PRL_2023}, MnF$_2$~\cite{BhowalPRX2024}, and MnTe~\cite{Lee2024_PRL,Mazin_PRBL_2023} etc. Subsequently, substantial development have been proposed in unconventional odd-parity noncollinear magnetic systems, termed as $p$-wave magnets~\cite{Linder_p_magnet_PRL,Nagae_2025,Hellenes2024pwavemagnets,Maeda_2024_p_magnet,Fukaya2025_pwave_magnet}. However, in contrast to AMs, $p$-wave magnets preserve TRS but break parity symmetry leading to spin-splitted band structure.
	
	Over the past few decades, superconductvity has always been a major area of research activity for realizing intricate phenomena. One of the fascinating phenomena is the superconducting proximity effect in mesoscopic systems where a non-vanishing supeconducting order is induced in a non-superconducitng material placed in close proximity to a bulk superconductor (SC)~\cite{Deutscher_1969,Bergeret_2005, Buzdin_2005}. Microscopically, the superconducting proximity effect emerges due to phase coherence between electron and hole wavefunctions even outside the superconducting region~\cite{Beenakker1997} leading to tunnelling of Cooper pairs into a nonsuperconducting material, thereby inducing superconducting correlations over some characteristic length scales. Proximity effects not only enable superconductivity in otherwise nonsuperconducting systems, but also strongly modify superconducting properties near interfaces. This gives rise to rich physics in a wide range of hybrid junctions-where the induced Cooper-pair symmetries have been extensively analyzed in the literature~\cite{Buzdin_2005,Fu_prox_2008, Burset_2015,Bena_2013,Khanna_2014,Sitthison_2014,Linder_2010,Black-Schaffer_1_2013,Black-Schaffer_2_2013, Paramita_2019,Paramita_2020, Eschrig_2015}.
	
	The detrimental interplay between magnetism and superconductivity has long been recognized as a fertile platform for realizing unconventional quantum phases. In this direction, focussing on AM-SC heterostructures~\cite{Fukaya_2025_Rev}, several theoretical investigation have been carried out which includes Andreev reflections~\cite{Papj_PRBL_2023,Sun_2023,Zhao_2025}, Josephson effects~\cite{Ouassou_PRL_2023,Beenakker_2023,Cheng_PRB_2024,pal_2025,Fukaya_2025,Cheng_2_PRB2024,Chakraborty2025b,debnath2025}, magnetoelectric responses~\cite{zyuzin2025,Hu_2025}, finite-momentum pairing~\cite{Zhang2024,Chakraborty_2024_1,Sumita_2025,Chakraborty2025a,rasmussen2025inherentmomentumdependentgapstructure}, and routes to engineer topological superconductivity~\cite{Ghorashi_PRL,Mondal_PRB_1,Li_PRBL_2024,Li_PRBL_2023,Zhu2023,Li2024,pal2025AM_SDE}. A central goal in this context is the realization of topological suprconductor (TSC), hosting Majorana modes as zero-energy quasiparticle excitations~\cite{Kitaev_2001,qi2011topological,Leijnse_2012,Alicea_2012,NayakRMP2008}. These neutral excitations are central to the proposals for fault-tolerant topological quantum computation owing to their non-Abelian braiding statistics~\cite{Ivanov2001,freedman2003topological,Stern2010}. While numerous engineered platforms have been proposed in one-dimension (1D) and $2$D~\cite{LutchynPRL2010,das2012zero,ramonaquado2017,Yazdani2013,Yazdani2015,Felix_analytics,Mondal_2023_Shiba,Mourik_Science2012,Beenakker2013search,Fu_prox_2008,Chatterjee_2024}--mostly rely on magnetic impurities or external magnetic fields, which tend to suppress superconductivity. On the other hand, AMs offer an appealing alternative in $2$D as they provide a source of broken TRS prevailing larger bulk superconducting gaps compared to the other platforms to realize TSC.

	\begin{figure}
		\centering
		\includegraphics[scale=1.2]{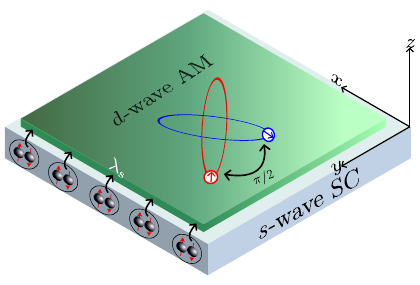} 
		\caption{Schematic illustration of our heterostructure comprised of a 2D $d$-wave AM layer placed on top of a three-dimensional bulk $s$-wave SC. Cooper-pair tunneling across the interface is described by the effective coupling strength $\lambda_{s}$.The two lobes of the spin-up and spin-down Fermi surfaces of the $d$-wave AM are schematically shown, and in this work we focus on the case where they are rotated by $90^{\circ}$ relative to each other.}
		\label{fig:fig_1}
	\end{figure} 
	
	Despite these advantages, most theoretical works so far treated the induced superconductivity phenomenologically~\cite{Ghorashi_PRL,Mondal_PRB_1,Maeda_2025,Li_PRBL_2023,Fukaya_2025_Rev}, by inserting a constant pairing term directly into the low-energy Hamiltonian of the non-superconducting system. While such models capture qualitative aspects, they neglect the microscopic details of tunneling and interface processes that are crucial for determining the induced pairing structure and its robustness for the realization of TSC and related phenomena.
	
	In this work, we present a microscopic theory of superconducting proximity effects in AM–SC heterostructures. Specifically, we investigate a system consisting of a  $2$D $d$-wave AM placed on top of a three-dimensional (3D) conventional $s$-wave SC. By integrating out the superconducting degrees of freedom~\cite{Bena_2013,Khanna_2014}, we derive an effective Hamiltonian and Green’s function therein for the AM layer that incorporates tunneling processes explicitly. This framework allows us to: (i) perform spectral analysis and (ii) characterize induced pairing symmetries, including mixed even- and odd-frequency channels. Additionally, we investigate the role of interfacial  Rashba spin–orbit coupling (RSOC) in generating odd-parity triplet components \cite{Shen_2024}, and analyse the emergence of weak and strong topological superconductivity in this hybrid setup. Using exact diagonalization and Chern number calculation~\cite{Fukui_2005}, we demonstrate that the AM–SC platform supports boundary Majorana edge modes (MEMs) characterized by appropriate topological invariants. Importantly, the tunneling amplitude appears as an experimentally tunable parameter that directly connects our microscopic description to realistic devices.
	
	The remainder of this paper is organized as follows. In Sec.~\ref{Sec:II_model}, we introduce our model Hamiltonian and derive the effective Green’s functions for the AM layer. In Sec.~\ref{Sec:III_dos}, we analyze the density of states (DOS) and induced pairing amplitudes in the AM layer. In Sec.~\ref{Sec:IV_Symm_RSOC}, we discuss the symmetry properties of the induced pairings in the presence of RSOC. Sec.~\ref{Sec:V_ED} is devoted to the discussion of exact diagonalization results for the DOS and benchmark them against the proximity induced effective model. In Sec.~\ref{Sec:topology}, the emergence of weak and strong topological superconducting phases is elaborated by computing topological invariants and comparing them with exact diagonalization spectra. Finally, in Sec.~\ref{Sec:conclusions}, we summarize our results and provide conclusions.

	\section{Model and  Method}\label{Sec:II_model}
	We begin by considering a heterostructure comprising of a 2D $d$-wave AM with $d_{x^2-y^2}$ magnetic order~\cite{Smejkal_PRX_1,Smejkal_PRX_2}, placed on top of a 3D ordinary bulk $s$-wave SC as schematically illustrated in Fig.\,\ref{fig:fig_1}. The model Hamiltonian for the AM layer can be written in the Bogoliubov de Gennes (BdG) basis as,
		\begin{equation}
		\begin{aligned}
			H_{AM} &= -\mu \sum_{\mathbf{r}} \Psi_{\mathbf{r}}^\dagger (\tau_z \sigma_0) \Psi_{\mathbf{r}} 
			- t \sum_{\langle \mathbf{r}, \mathbf{r}' \rangle} \Psi_{\mathbf{r}}^\dagger (\tau_z \sigma_0) \Psi_{\mathbf{r}'} \\
			&\quad + J_a \sum_{\mathbf{r}} \Psi_{\mathbf{r}}^\dagger \tau_0 \sigma_z \left( \Psi_{\mathbf{r}+\hat{x}} - \Psi_{\mathbf{r}+\hat{y}} \right) + \text{H.c.}\ ,
		\end{aligned}
		\label{eq:H_AM}
	\end{equation}
	where, $\Psi_{\mathbf{r}} = \begin{pmatrix} \psi_{\uparrow}, \psi_{\downarrow}, \psi_{\downarrow}^\dagger, -\psi_{\uparrow}^\dagger \end{pmatrix}_{\mathbf{r}}^T$ is the Nambu spinor in the BdG basis with $ \psi_{ \mathbf{r}s}~(\psi_{\mathbf{r}s}^\dagger)$ correponds to the electron annhilation (creation) operator at the site $\mathbf{r}=(x,y)$ with spin $s=(\uparrow,\downarrow)$ (actually pseudospin in case of AM ~\cite{Chakraborty2025a}) in the altermagnetic layer. Here, \(\sigma\) and \(\tau\) represent Pauli matrices in spin and particle-hole space, respectively and \(\mu\), \(t\) and \(J_a\)  denote the chemical potential, nearest-neighbor hopping amplitudes, and AM exchange energy respectively.
	

	The model Hamiltonian for the 3D $s$-wave SC reads as,
	%
	%
	\begin{equation}
		\begin{aligned}
			H_{SC} &= -\mu_{\text{sc}} \sum_{\mathbf{R}, \sigma} \Phi_{\mathbf{R}, \sigma}^\dagger \Phi_{\mathbf{R}, \sigma} 
			- t_{\text{sc}} \sum_{\langle \mathbf{R}, \mathbf{R}' \rangle, \sigma} \Phi_{\mathbf{R}, \sigma}^\dagger \Phi_{\mathbf{R}', \sigma} \\
			&\quad + \sum_{\mathbf{R}} \left( \Delta\, \Phi_{\mathbf{R}, \uparrow}^\dagger \Phi_{\mathbf{R}, \downarrow}^\dagger + \text{H.c.} \right)\ ,
		\end{aligned}
		\label{eq:H_SC}
	\end{equation}
	where, $\Phi_{\mathbf{R}}~(\Phi_{\mathbf{R}}^\dagger)$ corresponds to the fermionic annihilation (creation) operator in the SC at site ${\mathbf{R}}=(x,y,z)$. The parameters \(\mu_{\text{sc}}\), \(t_{\text{sc}}\)  and \(\Delta\) denote the chemical potential in the SC, nearest neighbour hopping amplitude and $s$-wave superconducting order, respectively.
	
	To microscopically incorporate the superconducting proximity effect, we consider a coupling between the top layer of the SC and the AM as,
	%
	\begin{equation}
		H_T = \tilde{t} \sum_{\mathbf{r}, \tau, \sigma} \, \psi_{\mathbf{r}, \tau, \sigma}^\dagger \Phi_{\mathbf{r} + \hat{z}, \sigma} + \text{H.c.}\ ,
		\label{eq:H_T}
	\end{equation}
	where, \(\mathbf{r}\) denotes the sites in the AM and \(\mathbf{r} + \hat{z}\) denotes the first layer of the SC. Here, \(\tilde{t}\) indicates the tunneling amplitude between the AM and SC.
	
	\begin{figure*}[t]
		\centering
		\includegraphics[width=1.0\textwidth]{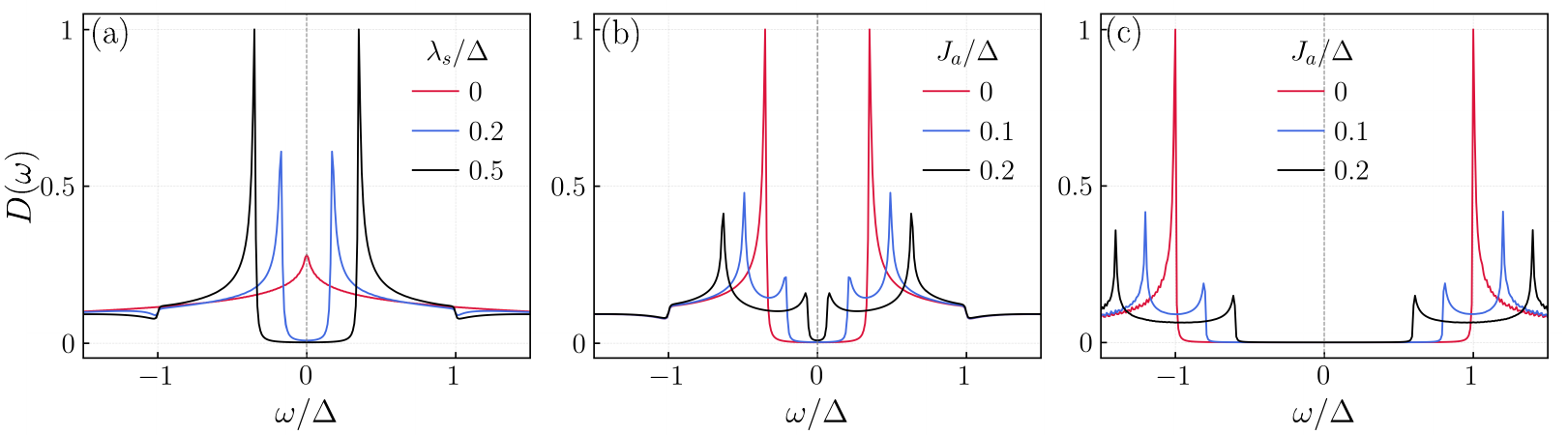}
		\caption{The total DOS $D(\omega)$ of the AM layer is shown for two cases: superconductivity is induced via the proximity effect [panels (a) and (b)] and superconducting pairing incorporated directly into the bulk Hamiltonian [panel (c)]. Panel (a) presents the results for $J_{a}=0$ with different values of $\lambda_{s}$, while panel (b) depicts the case for $\lambda_{s}=0.5\Delta$ with varying $J_{a}$. On the other hand, panel (c) illustrates the bulk SC case for different values of $J_{a}$. 
		The other model parameters used are $\mu=0$ and $\Delta=0.3t$.}
		\label{fig:fig_2}
	\end{figure*}    
	Since the Hamiltonian in the SC (see Eq.\,\eqref{eq:H_SC}) is quadratic in the fermionic degrees of freedom, we integrate it out to obtain an effective action for the AM. This allows us to explicitly focus on the altermagnetic layer as the physical subsystem of interest, with the superconducting degrees of freedom participating only through a self-energy term that encodes proximity-induced superconducting correlations in the AM. Following the procedure for computing the self-energy as outlined in~Refs.~\cite{Khanna_2014,Bena_2013,Cuevas_1996}, we decouple the AM and SC  degrees of freedom and calculate the Green's function \(G(\omega)\) for the altermagnetic layer (a sketch of the derivation is provided in Appendix~\ref{app:self_energy}) as,
	\begin{equation}
		G(\omega, k_x, k_y) = \left[ (\omega + i\eta)\, I - H_{AM}(k_x, k_y) - \Sigma(\omega) \right]^{-1}\ ,
		\label{eq:Green_function}
	\end{equation}
	where,
	\begin{equation}
		\begin{aligned}
			H_{AM}(k_x,k_y)
			&= \big[-\mu-2t(\cos k_x + \cos k_y)\big]\,\tau_z\sigma_0 \\
			&\quad + J_a\big(\cos k_x - \cos k_y\big)\,\tau_0\sigma_z\ ,
		\end{aligned}
		\label{eq:H_AM_1}
	\end{equation}
	is the momentum space Hamiltonian corresponding to the AM. Also, the
	self-energy due to the coupling with the SC can be written as
	\begin{equation}
		\Sigma(\omega) = \frac{\lambda_s}{\sqrt{\Delta^2 - (\omega - i\eta)^2}} 
		\left( \omega\, I + \Delta\, \tau_x \sigma_0 \right).
		\label{eq:SelfEnergy}
	\end{equation}
Here, $\lambda_s = \pi N(0) |\tilde{t}|^2$ characterizes the effective coupling strength to the SC, with $N(0)$ denoting the normal-state DOS at the Fermi level of the SC. In Eq.~(\ref{eq:SelfEnergy}), $\eta$ refers to a positive infinitesimal number that ensures the causality of the Green's function. The coupling strength between the AM and SC layer is primarily controlled by the $N(0)$, and $\tilde{t}$. Note that, $N(0)$ depends on the model parameters of the SC, such as hopping amplitude and Fermi energy of the SC, thus cannot take arbitrary values. In addition, the strength of $\tilde{t}$ depends on the interfacial properties of the AM-SC layer which we assume to be weak compared to other model parameters $t$, $\Delta$. For this reasons, the strength of $\lambda_s$ is constrained by $N(0)$ and $\tilde{t}$, and cannot take arbitrary large values within the weak-coupling regime. Additionally, it seems that for $\lambda_s>\Delta$, the proximity-induced gap becomes larger than the superconducting gap of the parent SC. However, physically such enhancement is not possible as our formalism is mainly valid in the weak coupling regime i.e., $\lambda_s<\Delta$. Thus, in particular, one must avoid parameter values for which $\lambda_s$ becomes larger than $\Delta$. To ensure that, all numerical results presented in our manuscript are restricted to the regime $\lambda_s \le 0.5\Delta$.

	
	
	\section{DOS and Pairing Amplitude}\label{Sec:III_dos}
	In this section, we discuss the DOS of the altermagnetic layer in the presence of superconducting pairing amplitudes induced by the $s$-wave SC as a result of superconducting proximity effect. We compute the DOS of the proximity induced altermagnetic layer using the expression,
	\begin{equation}
		D(\omega) = -\frac{1}{\pi} \int_{\mathrm{BZ}} \frac{d^2 \mathbf{k}}{(2\pi)^2} \, 
		\mathrm{Im} \left[ \mathrm{Tr} \, G(\omega, k_x, k_y) \right],
		\label{eq:DOS}
	\end{equation}
	where, the trace is performed over both spin- and particle-hole degrees of freedom, and the integral is computed over the Brillouin zone (BZ). Here $\textbf{k}=(k_x,k_y)$ denotes momentum components along $x$ and $y$-directions. The variation of DOS, $D(\omega)$ as a function of energy is presented in Fig.\,\ref{fig:fig_2} for various circumstances. In Fig.~\ref{fig:fig_2}(a), we examine the case in absence of AM exchange interaction (\(J_a = 0\)) to analyse the sole effect of superconducting proximity effect. Moreover, in the absence of coupling with the SC (\(\lambda_s = 0\)), the DOS exhibits the usual feature 
	of normal 2D tight-binding Hamiltonian where the pronounced peak at zero energy corresponds to the van Hove singularity. 
	Introducing a finite AM-SC coupling \(\lambda_s\) leads to the formation of proximity induced superconducting gap in the DOS. As $\lambda_s$ increases, the induced superconducting proximity in the AM layer becomes stronger, and a larger gap appears, indicating enhanced superconducting correlations. Then, Fig.~\ref{fig:fig_2}(b) illustrates the evolution of the DOS as one varies AM exchange energy \(J_a\) at a fixed coupling strength \(\lambda_s = 0.5\Delta\), where a SC gap is already induced even when $J_{a}=0$.  As \(J_a\) increases, the proximity-induced gap is gradually suppressed, signaling a competition between superconductivity and the AM exchange field, and eventually gets destroyed at a critical value \(J_a \sim \lambda_s/2\). This limit has been derived in Appendix~\ref{Appendix_limit}.
	In addition to the SC coherence peaks, finite value of \(J_a\) leads to spin-splitting of the quasiparticle bands which is manifested as the splitting of the coherence peaks (see Fig.~\ref{fig:fig_2}(b)). In Fig.~\ref{fig:fig_2}(c), we present the DOS for the case where superconductivity is incorporated directly into the AM layer by adding a static pairing term to its Hamiltonian, without explicitly considering the microscopic tunneling mechanism  (the expression for the corresponding phenomenological Hamiltonian is given in Appendix~\ref{App.Phenomeno}). Although the resulting DOS exhibits qualitative characteristics similar to those in Fig.~\ref{fig:fig_2}(b), the two approaches differ significantly from physical point of view.
	
	We now turn our focus to analyse the proximity induced superconducting pairing amplitudes in the altermagnetic layer. We begin by rewriting the Nambu-Gorkov Green's function, defined in Eq.\,\eqref{eq:Green_function} in terms of it's components as,
	\begin{equation}
		G_{\mathbf{k}}(\omega) = 
		\begin{pmatrix}
			g(\omega, \mathbf{k}) & \bar{\Delta}(\omega, \mathbf{k}) \\
			\bar{\Delta}^\dagger(\omega, \mathbf{k}) & g'(\omega, \mathbf{k})
		\end{pmatrix}\ .
		\label{eq:G_matrix}
	\end{equation}
	This is a \(4 \times 4\) matrix in Nambu space, where the off-diagonal components \(\bar{\Delta}(\mathbf{k}, \omega)\) and \(\bar{\Delta}^\dagger(\mathbf{k}, \omega)\) denote the anomalous Green's functions, which are \(2 \times 2\) matrices in spin space and are given by,
	\begin{equation}
		\bar{\Delta}(\mathbf{k}, \omega) =
		\begin{pmatrix}
			\Delta_{\uparrow\downarrow}(\mathbf{k}, \omega) & -\Delta_{\uparrow\uparrow}(\mathbf{k}, \omega) \\
			\Delta_{\downarrow\downarrow}(\mathbf{k}, \omega) & -\Delta_{\downarrow\uparrow}(\mathbf{k}, \omega)
		\end{pmatrix}\ .
		\label{eq:anomalous_GF}
	\end{equation}
	
	The anomalous Green’s function encodes the SC correlations in the AM system. To identify the nature of these correlations, it is necessary to analyze the symmetry properties of the pair amplitude \( \Delta_{\sigma\sigma'}(\mathbf{k}, \omega) \).
	This classification is guided by the antisymmetry constraint imposed by the Fermi–Dirac statistics on the pair amplitude under the total exchange of the quantum numbers~\cite{Tanaka_2012, Cayao2020, Maeda_2025, Dutta2021,Fukaya_2025_Rev} as
	\begin{equation}
		\bar{\Delta}_{\sigma \sigma'}(\mathbf{k}, \omega) = -\bar{\Delta}_{\sigma' \sigma}(-\mathbf{k}, -\omega)\ .
		\label{eq:antisymmetry}
	\end{equation}
	The pair amplitude may exhibit either even or odd behavior under the individual exchange of spin indices, momentum, or frequency. However, the total exchange of all quantum numbers must result in a minus sign, as required by Eq.~\eqref{eq:antisymmetry}.

	Without loss of generality, the spin structure of the pairing amplitude can be decomposed as follows~\cite{Sigrist_1991},
	\begin{equation}
		\bar{\Delta}(\mathbf{k}) = \psi(\mathbf{k})+ \mathbf{d}(\mathbf{k}) \cdot \boldsymbol{\sigma}\ ,
		\label{eq:decouple}
	\end{equation}
	where, \(\psi(\mathbf{k}, \omega)\) corresponds to the spin-singlet Cooper pairs, and the vector \(\mathbf{d}(\mathbf{k}, \omega)\) represents the spin-triplet Cooper pairs. They can be written as
	\begin{equation}
		\begin{aligned}
			\psi(\mathbf{k}, \omega) &= \tfrac{1}{2}\left[\Delta_{\uparrow\downarrow}(\mathbf{k}, \omega) - \Delta_{\downarrow\uparrow}(\mathbf{k}, \omega)\right]\ , \\
			d_x(\mathbf{k}, \omega) &= \tfrac{1}{2}\left[-\Delta_{\uparrow\uparrow}(\mathbf{k}, \omega) + \Delta_{\downarrow\downarrow}(\mathbf{k}, \omega)\right]\ , \\
			d_y(\mathbf{k}, \omega) &= \tfrac{1}{2i}\left[\Delta_{\uparrow\uparrow}(\mathbf{k}, \omega) + \Delta_{\downarrow\downarrow}(\mathbf{k}, \omega)\right]\ , \\
			d_z(\mathbf{k}, \omega) &= \tfrac{1}{2}\left[\Delta_{\uparrow\downarrow}(\mathbf{k}, \omega) + \Delta_{\downarrow\uparrow}(\mathbf{k}, \omega)\right]\ .
		\end{aligned}
		\label{eq:pairing-decomposition}
	\end{equation}

	Before proceeding further, we classify the possible symmetries of the superconducting correlations. Based on the antisymmetry condition imposed by Eq.~(\ref{eq:antisymmetry}), the pairing amplitudes fall into four distinct symmetry classes:  
	(i) even-frequency spin-singlet even parity (ESE),  
	(ii) even-frequency spin-triplet odd parity (ETO),  
	(iii) odd-frequency spin-singlet odd parity (OSO), and  
	(iv) odd-frequency spin-triplet even parity (OTE).  
	
	Although all four symmetry classes are, in principle, allowed, their presence and relative magnitudes depend sensitively on the coupling strength and the characteristics of the system. For visualization purposes, we define a quantity \(\phi_\beta(\omega)\), which captures the contribution arising from the singlet or triplet pair amplitude at a given frequency~\cite{Maeda_2025},
	\begin{equation}
		\phi_\beta(\omega) = \int_{\mathrm{BZ}} d^2 k \, d_\beta(\omega, \mathbf{k})\, d_\beta^\dagger(\omega, \mathbf{k})\ ,
		\label{eq:phi_alpha}
	\end{equation}
	where, \(\beta = s, t\), with \(d_s = \psi\) representing the singlet component and \(d_t = \{d_x, d_y, d_z\}\) denoting the triplet components.
	
	For our model, the singlet and triplet components can be obtained analytically as 
	\begin{equation}
		\begin{aligned}
			\psi(\omega, \mathbf{k}) &= 
			\frac{q(\omega) \chi(\omega, \mathbf{k})}{\chi(\omega, \mathbf{k})^2 - 4 h_{AM}(\mathbf{k})^2 p(\omega)^2}\ , \\
			d_z(\omega, \mathbf{k}) &= 
			\frac{2 h_{AM}(\mathbf{k}) p(\omega) q(\omega)}{\chi(\omega, \mathbf{k})^2 - 4 h_{AM}(\mathbf{k})^2 p(\omega)^2}\ , \\
			& d_x(\omega, \mathbf{k})  = d_y(\omega, \mathbf{k}) = 0.	 
		\end{aligned}
		\label{eq:pair_components}
	\end{equation}
	where, the auxiliary functions are defined as
	\begin{equation}
		\begin{aligned}
			&\chi(\omega, \mathbf{k}) = h_{AM}^2 - h_t^2 + p^2 - q^2, \\
			&h_t(\mathbf{k}) = -2t(\cos k_x + \cos k_y) - \mu, \\
			&h_{AM}(\mathbf{k}) = J_a(\cos k_x - \cos k_y), \\
			&p(\omega) = \omega + i\eta + \frac{\lambda_s \omega}{\sqrt{\Delta^2 - (\omega - i\eta)^2}}, \\
			&q(\omega) = \frac{\lambda_s \Delta}{\sqrt{\Delta^2 - (\omega - i\eta)^2}}.
		\end{aligned}
		\label{eq:Auxiliary_terms}
	\end{equation}
		\begin{figure}[t]
		\centering
		\includegraphics[width=\columnwidth]{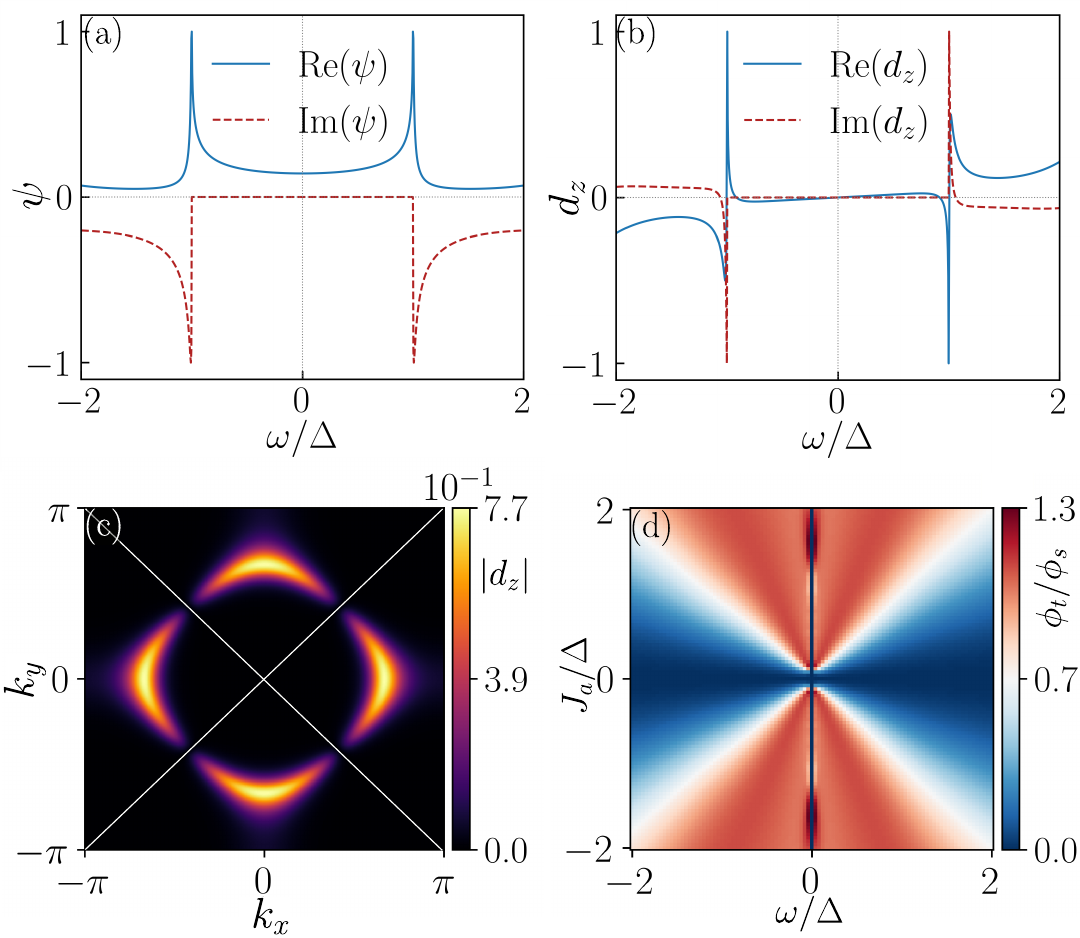}
		\caption{In panels (a) and (b), we display the real and imaginary parts of $\psi(\omega, \mathbf{k})$ and $d_z(\omega, \mathbf{k})$, respectively, as functions of $\omega$ for fixed momentum values $(k_x, k_y) = (1, 0.3)$ (in units of the inverse lattice constant). Panel (c) showcases the absolute value of the triplet amplitude $|d_z|$ in the $k_x$–$k_y$ plane at $\omega = 0.5\Delta$, where the white diagonal lines indicate nodes at which $|d_z|$ vanishes below a numerical tolerance of $10^{-18}$. Panel (d) depicts the ratio of the integrated squared magnitudes of OTE to ESE amplitudes in the $J_a$–$\omega$ plane. Panels (a)–(c) are obtained for $J_a = 0.1\Delta$. The other model parameters are chosen as $\mu = 0$, $\Delta = 0.3t$, and $\lambda_s = 0.5\Delta$.}
		\label{fig:fig_3}
	\end{figure}
	
	In order to carry out the calculation of frequency-resolved pairing amplitudes, we use the retarded Green’s functions for $\omega < 0$ and the advanced Green’s functions for $\omega > 0$~\cite{Parhizgar2020}. By inspection of Eq.~\eqref{eq:pair_components}, the spin-singlet component 
	\(\psi(\mathbf{k}, \omega)\) is an even function of both frequency and momentum. 
	This follows from the fact that \(\chi(\omega, \mathbf{k}) = \chi(\omega, -\mathbf{k})\), 
	and is further corroborated by the results presented in Fig.~\ref{fig:fig_3}(a), 
	where both the real and imaginary parts of \(\psi(\mathbf{k}, \omega)\) remain even as a function of \(\omega\). 
	We therefore identify \(\psi\) as belonging to the ESE class. 
	In contrast, from Eq.~\eqref{eq:pair_components}, it is clear that the momentum dependence of the spin-triplet component \(d_z(\mathbf{k}, \omega)\) arises from \(\chi(\omega, \mathbf{k})\) and \(h_{\mathrm{AM}}(\mathbf{k})\). Note that, both of them are even functions of \(\mathbf{k}\). As shown in Fig.~\ref{fig:fig_3}(b), both the real and imaginary parts of \(d_z(\mathbf{k}, \omega)\) are odd functions of \(\omega\). Hence, \(d_z(\mathbf{k}, \omega)\) belongs to the OTE class, consistent with the antisymmetry condition of Eq.~\eqref{eq:antisymmetry}. 
	Importantly, \(d_z(\mathbf{k}, \omega)\) originates from the \(d\)-wave altermagnetic field \(h_{\mathrm{AM}}(\mathbf{k})\), which acts as a spin-mixing mechanism enabling triplet correlations in the AM layer (a similar pairing symmetry analysis for the newly proposed $p$-wave magnet is presented in Appendix~\ref{Sec_p_wave}). To illustrate these features, Fig.~\ref{fig:fig_3}(c) displays the absolute value of the frequency resolved OTE amplitude \(|d_z|\). Four nodal lines appear along the diagonals, reflecting the underlying \(d\)-wave structure of the AM. Such nodal-line structure originates from the characteristic momentum-dependent spin splitting of AMs and constitutes a qualitative distinction from conventional ferromagnetic systems, where such nodal features are generically absent. Thus, while ESE pairing is directly inherited from the parent $s$-wave SC, 
	OTE pairing appears from the interplay between the $d$-wave AM and the spin-singlet SC. The relative weight of the induced triplet component is quantified in 
	Fig.~\ref{fig:fig_3}(d), which manifests the ratio of the integrated squared magnitudes 
	of OTE and ESE amplitudes across the $(J_a,\omega)$ plane. 
	A finite ratio over a broad parameter range demonstrates the coexistence of both the components. Notably, OTE pairing becomes dominant when $J_a \sim \lambda_s/2$, as indicated by the red regions. Although symmetry analysis categorizes $d_z$ as OTE, i.e., even in parity, a finite $p$-wave–like structure usually requires odd parity pairing. 
	In the following, we explore the role of RSOC in the AM layer 
	as a natural mechanism that can promote such momentum-dependent triplet terms. 
	\begin{figure*}[t]
		\centering
		\includegraphics[width=\textwidth]{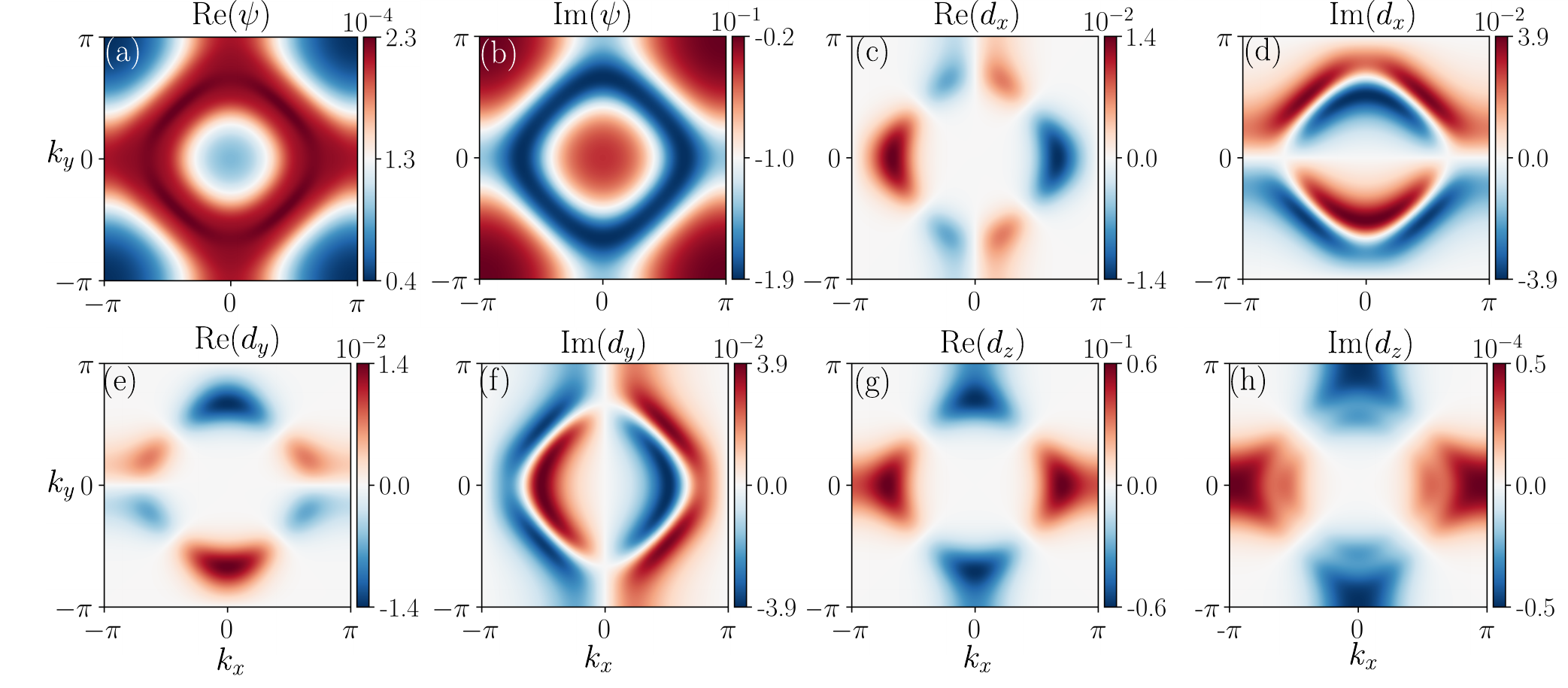}
		\caption{Momentum-resolved pairing amplitudes are depicted in the $k_{x}$-$k_{y}$ plane choosing $\omega = 0.5\Delta$. Panels (a) and (b) show the 
		real and imaginary parts of the singlet component $\psi(\mathbf{k})$. On the other hand, panels (c)–(h) display the real and imaginary parts of the triplet components $d_x(\mathbf{k})$, $d_y(\mathbf{k})$, and $d_z(\mathbf{k})$. The finite in-plane triplet components $d_x$, $d_y$ arise in the presence of RSOC [$\alpha=0.3\Delta$]. The other model parameters used are $J_a=0.1\Delta$, $\mu=0$, $\Delta=0.3 t$, $\lambda_s=0.5 \Delta$.}
		\label{fig:fig_4}
	\end{figure*}

	\section{Pairing Symmetry with RSOC}\label{Sec:IV_Symm_RSOC}
	\begin{figure}
		\centering
		\includegraphics[width=\columnwidth]{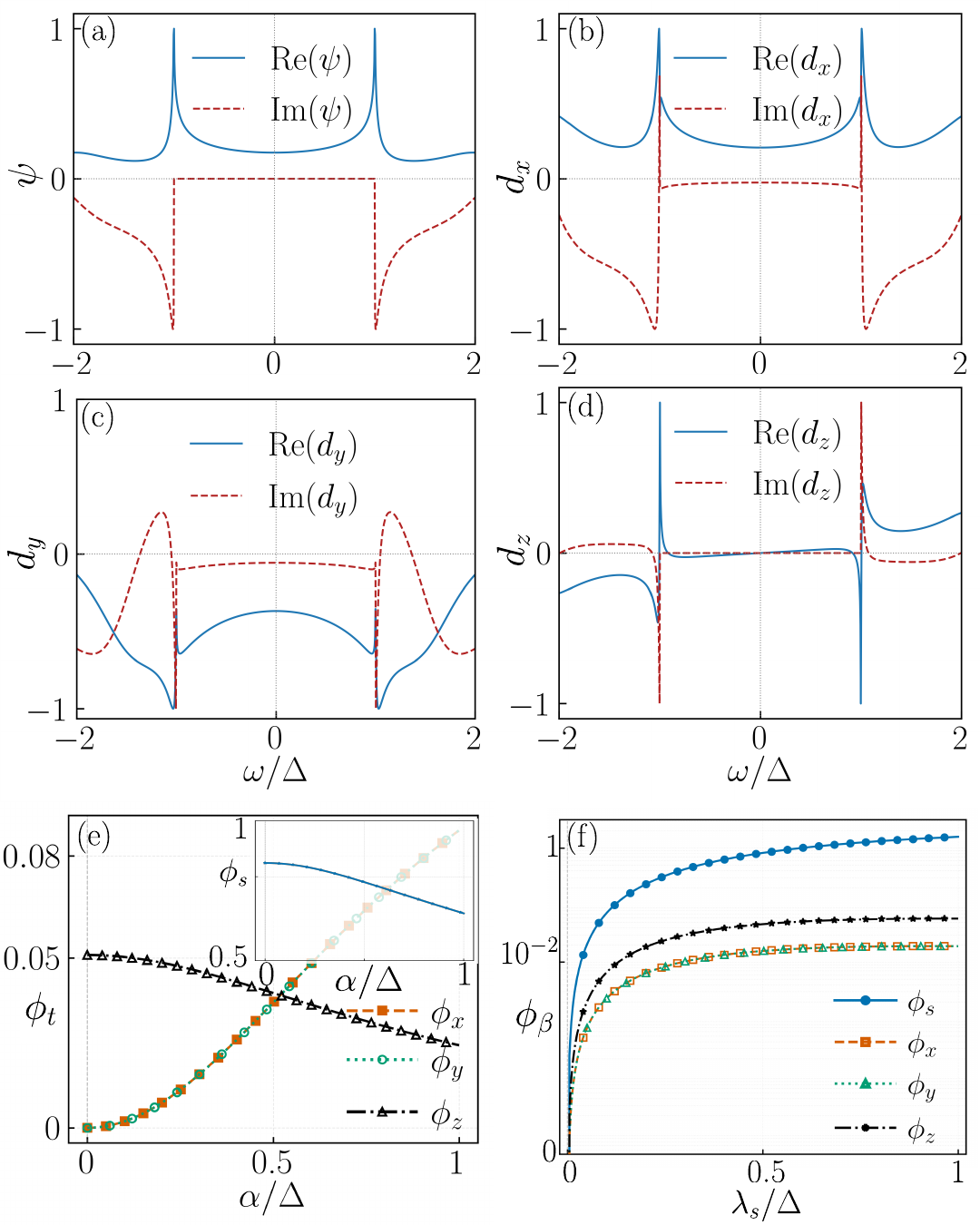}
		\caption{
			Panels (a)–(d) demonstrate the frequency dependence of the pairing amplitudes at an arbitrary fixed momentum point $(k_x, k_y) = (1, 0.4)$ (in units of the inverse lattice constant) in the Brillouin zone. Panel (a) displays the real and imaginary parts of the singlet component, panels (b) and (c) present the in-plane triplet components $d_x$, $d_y$, and panel (d) illustrates the transverse triplet component $d_z$. In panel (e), we exhibit the absolute magnitudes of the triplet components as a function of the Rashba SOC strength $\alpha$ (the inset depicts the singlet component), while panel (f) presents the absolute magnitudes of both singlet and triplet components as a function of the AM-SC coupling strength $\lambda_s$, plotted on a logarithmic scale. The other model parameters are chosen as $J_a=0.1\Delta$, $\mu=0$, $\Delta=0.3 t$, $\lambda_s=0.5 \Delta$.
		}
		\label{fig:fig_5}
	\end{figure}
		From the previous analysis, we find that a pure \(d\)-wave AM does not generate odd-parity (\(p\)-wave) triplet correlations via the proximity effect. However, the incorporation of an additional spin-mixing mechanism such as RSOC in the AM layer gives rise to odd-parity triplet
		component.  The combined AM+RSOC Hamiltonian in momentum space reads as 
		\begin{align}
			H_{\mathrm{AM+RSOC}}(\mathbf{k}) 
			& = h_t(\textbf{k})\,\tau_z\sigma_0 
			+h_{AM}(\textbf{k})\,\tau_0\sigma_z \nonumber\\ 
			&+ 2\alpha(\sin k_y\,\tau_z\sigma_x - \sin k_x\,\tau_z\sigma_y)\ ,
			\label{eq:H_AM_RSOC}
		\end{align}
	where, \(\alpha\) is the RSOC strength. Note that, the RSOC term in Eq.~(\ref{eq:H_AM_RSOC}) is incorporated assuming an interfacial layer between the AM and SC, not intrinsic to AM which would compromise the altermagnetic classification. However, for simplicity, we model the combined effect of AM and RSOC by the effective tight-binding Hamiltonian $H_{\mathrm{AM+RSOC}}(\mathbf{k})$ instead of considering a microscopic tunneling between the AM and RSOC layer, or, considering the presence of RSOC in the SC layer itself~\cite{Heinsdorf2026}. Following the previous section, we obtain the analytic expressions for the singlet and triplet components of the pairing amplitudes as
	\begin{equation}
		\begin{aligned}
			\psi(\omega, \mathbf{k}) &= \frac{q \left( h_{\mathrm{rx}}^2 + h_{\mathrm{ry}}^2 + h_t^2 - h_{\mathrm{AM}}^2 - p^2 + q^2 \right)}{\Gamma(\textbf{k},\omega)}\ , \\
			&d_x(\omega, \mathbf{k}) = \frac{2q \left( i h_{\mathrm{rx}} h_{\mathrm{AM}} - h_{\mathrm{ry}} h_t \right)}{\Gamma(\textbf{k},\omega)}\ , \\
			&d_y(\omega, \mathbf{k}) = \frac{2q \left( h_{\mathrm{rx}} h_t + i h_{\mathrm{ry}} h_{\mathrm{AM}} \right)}{\Gamma(\textbf{k},\omega)}\ , \\
			&d_z(\omega, \mathbf{k}) = -\frac{2p q h_{\mathrm{AM}}}{\Gamma(\textbf{k},\omega)}\ .
		\end{aligned}
		\label{eq:pair_components_rsoc}
	\end{equation}
	where, the denominator is given by 
	\begin{widetext}
		\begin{multline}
			\Gamma = h_{\mathrm{AM}}^4 + h_{\mathrm{rx}}^4 + h_{\mathrm{ry}}^4 - 2 h_{\mathrm{ry}}^2 h_t^2 + h_t^4 
			- 2 h_{\mathrm{ry}}^2 p^2 - 2 h_t^2 p^2 + p^4 + 2 h_{\mathrm{ry}}^2 q^2 + 2 h_t^2 q^2 - 2 p^2 q^2 + q^4 \\
			+ 2 h_{\mathrm{AM}}^2 ( h_{\mathrm{rx}}^2 + h_{\mathrm{ry}}^2 - h_t^2 - p^2 - q^2 )
			+ 2 h_{\mathrm{rx}}^2 ( h_{\mathrm{ry}}^2 - h_t^2 - p^2 + q^2 )\ ,
			\label{eq:Gamma}
		\end{multline}
	\end{widetext}
	with $h_{\mathrm{rx}}(\mathbf{k}) = 2\alpha \sin k_y$, $h_{\mathrm{ry}}(\mathbf{k}) = 2\alpha \sin k_x$, $p(\omega)$, $q(\omega)$, $h_{\mathrm{AM}}$ and $h_t$ as defined in Eq.~\eqref{eq:Auxiliary_terms}.
	Since the denominator $\Gamma(\textbf{k},\omega)$ is an even function of momentum, the parity of each pairing component is determined solely by its numerator. Both the singlet $\psi$ and out-of-plane triplet $d_z$ components are even in momentum, while the in-plane triplet components $d_x$ and $d_y$ are odd in momentum due to the terms arising from the RSOC $h_{\mathrm{rx}}$ and $h_{\mathrm{ry}}$ respectively. 
	This momentum dependence is clearly visible in Fig.~\ref{fig:fig_4}, which highlights the real and imaginary parts of the frequency-resolved singlet and triplet components in the \(k_x\)–\(k_y\) plane. From Figs.~\ref{fig:fig_4}(a), (b), (g), and (h), it is evident that the singlet \(\psi\) and the out-of-plane triplet component \(d_z\) exhibit even parity in momentum space, whereas Figs.~\ref{fig:fig_4}(c)–(f) indicate that the in-plane triplet components \(d_x\) and \(d_y\) display odd parity. Notably, these odd-parity in-plane components are absent but appear once RSOC is introduced.
	To verify their symmetry in frequency space, in Figs.~\ref{fig:fig_5}(a)–(d) we display the real and imaginary parts of the momentum-resolved components \(\psi\), \(d_x\), \(d_y\), and \(d_z\) as functions of \(\omega\). From their features, it is evident that $\psi$ and the in-plane triplets ($d_x, d_y$) are even in frequency, while the out-of-plane component $d_z$ turns out to be odd in frequency. Note that, every point in the BZ possesses the same symmetry properties, although they can yield quantitatively different figures. Therefore, the antisymmetry condition of Eq.~\eqref{eq:antisymmetry} is satisfied: 
	\begin{itemize}
		\item $\psi$ belongs to the ESE class, 
		\item $d_x$ and $d_y$ belong to the ETO class, 
		\item $d_z$ belongs to the OTE class. 
	\end{itemize}
	Hence, the presence of RSOC enables the generation of in-plane $p$-wave superconducting correlations.

Finally, examining the evolution of the absolute magnitudes of the pairing components as functions of the RSOC strength \(\alpha\) and the coupling parameter \(\lambda_s\), we observe distinct trends as illustrated in Figs.~\ref{fig:fig_5}(e) and (f). In Fig.~\ref{fig:fig_5}(e), enhancement of the RSOC strength \(\alpha\) suppresses both the singlet \(\psi\) (see the inset of Fig.~\ref{fig:fig_5}(e)) and the out-of-plane triplet \(d_z\) components, while simultaneously enhancing the in-plane triplet components \(d_x\) and \(d_y\). On the other hand, it is evident from Fig.~\ref{fig:fig_5}(f) that increasing \(\lambda_s\) enhances the magnitudes of all pairing components, with the singlet \(\psi\) grows more rapidly than the triplets as the bulk SC is $s$-wave type. This contrasting behavior highlights the competing roles of RSOC and coupling of the AM layer to the bulk SC, where stronger RSOC drives the system towards dominant odd-parity \(p\)-wave pairing amplitudes.
	
\begin{figure}[t]
	\centering
	\includegraphics[width=\linewidth]{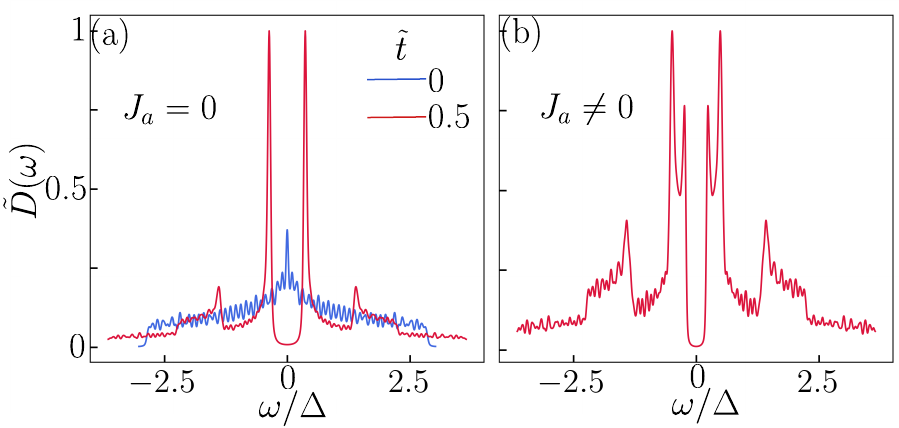}
	\caption{
		Numerical ED results for the DOS of the AM layer is presented. Panel (a) exhibits the case $J_a = 0$ for two different values of $\tilde{t}$. 
		Panel (b) presents the same for $J_a = \tilde{t} = 0.5\Delta$. We choose the other model parameters as $t_{\text{SC}} = t$, $\mu_{\text{SC}} = 0$, $\mu=0$ and $\Delta = 0.7t_{\text{SC}}$.
	}
	\label{fig:fig_6}
\end{figure}
	\begin{table}
		\centering
		\small
		\renewcommand{\arraystretch}{1.2}
		\caption{Classification of singlet and triplet pairing symmetries 
			in $d$-wave AM/$s$-wave SC hybrid system with and without RSOC.}
		\vspace{0.2cm}
		\begin{tabular}{|c|c|c|c|c|}
			\hline
			System & Pairing & Frequency & Parity & Symmetry \\
			\hline
			\multirow{2}{*}{\makecell{$d$-wave AM \\ + $s$-wave SC}} 
			& Singlet ($\psi$) & Even & Even & ESE \\
			\cline{2-5}
			& Triplet ($d_z$) & Odd & Even & OTE \\
			\hline
			\multirow{3}{*}{\makecell{$d$-wave AM \\ + $s$-wave SC \\ + RSOC}} 
			& Singlet ($\psi$) & Even & Even & ESE \\
			\cline{2-5}
			& Triplet ($d_z$) & Odd & Even & OTE \\
			\cline{2-5}
			& Triplet ($d_x,d_y$) & Even & Odd & ETO \\
			\hline
		\end{tabular} \label{Table:I}
	\end{table}
	

	\section{Comparison with Exact Diagonalization Results}\label{Sec:V_ED}

	To validate our preceding Green's function based analysis, we here perform exact diagonalization (ED) of the full 2D tight-binding Hamiltonian for the coupled system. The Hamiltonian, constructed in the combined Nambu basis that includes both the AM and SC degrees of freedom, can be written as
	\begin{equation}
		\mathcal{H}(\mathbf{k}) =
		\begin{pmatrix}
			H_{\mathrm{AM}}(\mathbf{k}) & H_{T}(\mathbf{k}) \\
			H_{T}^\dagger(\mathbf{k})  & H_{\mathrm{SC}}(\mathbf{k})
		\end{pmatrix}\ .
		\label{eq:full_hamiltonian}
	\end{equation}
	where, \(H_{\mathrm{AM}}\) and \(H_{\mathrm{SC}}\) correspond to the Hamiltonians of the isolated AM and SC subsystems, respectively, and \(H_T\) encodes the tunneling between them. The total Hamiltonian acts on a Hilbert space of dimension \(4\times N_z \times N_x \times N_y\), accounting for both spin and particle-hole degrees of freedom in each subsystem and $N_z$ is the number of SC layers. Diagonalization of $\mathcal{H}(\mathbf{k})$ yields eigenvalues $E_n$ and eigenvectors $\ket{\Xi_n}$. The local density of states (LDOS) in the AM region is computed by projecting these eigenstates onto the AM subspace. The projection is implemented using the operator $\Pi_{\mathrm{AM}} = \sum_{\mathbf{r} \in \mathrm{AM}} \sum_{\sigma,\tau} \left( \ket{\mathbf{r},\sigma,\tau}\bra{\mathbf{r},\sigma,\tau}\right)$. Here, the operator \(\Pi_{\mathrm{AM}}\) satisfies \(\Pi_{\mathrm{AM}}^2 = \Pi_{\mathrm{AM}}\) and isolates the AM layer’s degrees of freedom from the total Nambu basis. In practice, it can be represented as a diagonal matrix with unit entries corresponding to the AM orbitals and zeros elsewhere. The LDOS is then given by:
\begin{align}
	\mathrm{LDOS}_{\mathrm{AM}}(\mathbf{r}, \omega)
	= & \sum_n \sum_{\sigma = \uparrow, \downarrow, \tau} 
	\left| \bra{\mathbf{r}, \sigma, \tau} \Pi_{\mathrm{AM}} \ket{\Xi_n} \right|^2 \nonumber\\
	& \times \left[ \frac{1}{\pi} 
	\frac{\eta}{(\omega - E_n)^2 + \eta^2} \right]\ ,
	\label{eq:ldos_am_tau}
\end{align}
where, $\eta$ is a small broadening parameter, and $\ket{\mathbf{r}, \sigma , \tau}$ represents state in position space in the AM region.
Therefore, the frequency-resolved DOS for the entire AM layer is then obtained by summing over all sites as
	\begin{equation}
		\mathrm{\tilde{D}(\omega)} = \sum_{\mathbf{r}} \mathrm{LDOS}_{\mathrm{AM}}(\mathbf{r},\omega)\ .
		\label{eq:dos_am}
	\end{equation}

Using Eq.~\eqref{eq:dos_am}, we calculate the DOS in three distinct parameter regimes as mentioned below,
	\begin{itemize}
		\item[(a)] When both the inter-layer hopping and the altermagnetic exchange field are absent (\ie $\tilde{t}=0$ and $J_a=0$), the AM layer exhibits a metallic DOS with a peak at zero-energy akin to 2D tight-binding model (see Fig.~\ref{fig:fig_6}(a)). This result is analogous to the $\lambda_s=0$ case shown in Fig.~\ref{fig:fig_2}(a).
		\item[(b)] Introducing a finite inter-layer hopping amplitude ($\tilde{t} \neq 0$) opens a superconducting gap around zero energy in the AM layer, indicative of proximity-induced superconductivity (see Fig.~\ref{fig:fig_6}(a)). This mimics the finite $\lambda_s$ behavior presented in Fig.~\ref{fig:fig_2}(a). Interestingly, an additional pair of coherence peaks with lower amplitude appears in Fig.\,\ref{fig:fig_6}(a). This corresponds to the coherence peaks of the parent SC and can appear in the AM as higher order process. These additional peaks are absent in Fig.\,\ref{fig:fig_2}(a) where the SC degrees of freedom are integrated out as described using the Green’s-function approach in the tunneling regime (see Appendix\,\ref{app:self_energy} for details).
		\item[(c)] Finally, activating the altermagnetic exchange ($J_a \neq 0$) leads to a slight suppression of the induced pairing gap and the emergence of two additional Zeeman-like peaks, a signature of broken time-reversal symmetry. This is depicted in Fig.~\ref{fig:fig_6}(b). Note that, this result closely resembles the finite-$J_a$ case presented in Fig.~\ref{fig:fig_2}(b).
	\end{itemize}
	The close agreement between the numerical ED results and those obtained from the effective Green's function approach provides a numerical validation of our analytical framework.
	
	\section{Emergence of Topological Superconducting Phases}\label{Sec:topology}
	
	\begin{figure}
		\centering
		\includegraphics[width=\linewidth]{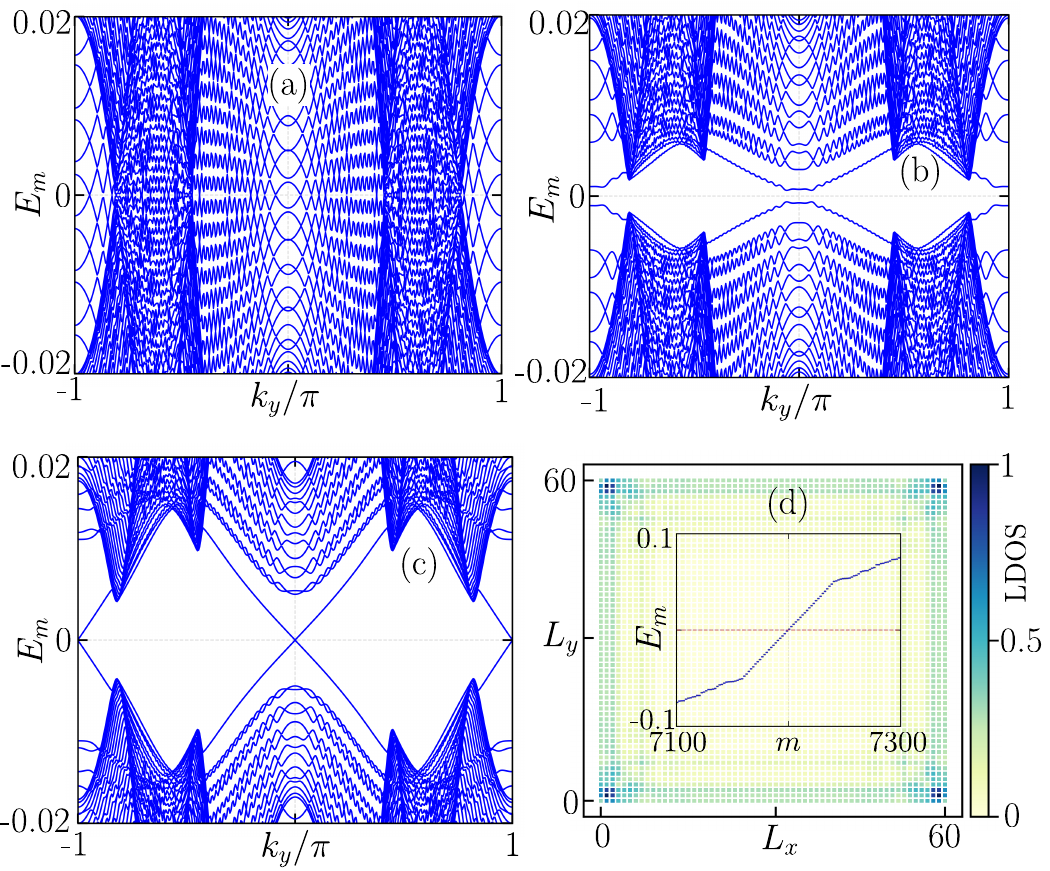}
		\caption{
			Panels (a)–(c) exhibit the edge spectra employing OBC along the $x$-axis and PBC along the $y$-axis for three different coupling strengths: $\lambda_s = 0$, $0.2 \Delta$, and $0.5 \Delta$. Panel (d) represents the site-resolved LDOS distribution at zero energy with OBC along both $x$ and $y$-directions. We consider a finite square lattice with $60 \times 60$ lattice sites at $\lambda_s = 0.5\Delta$. The inset displays a zoomed-in view of the quasi-energy spectrum near the zero-energy. The other model parameters are used as $t = 1$, $\mu = 0.4t$, $\Delta = 0.3t$, $b = 1$, $\alpha = 0.3t$, and $J_a = 0.3t$.
		}
		\label{fig:7edge_spectra_ldos}
	\end{figure}
	We have already seen that the inclusion of RSOC leads to the generation of odd-parity $p$-wave pairing amplitudes, $d_x$ and $d_y$ in the ETO class 
	(see Table\,\ref{Table:I}), and therefore we  expect the emergence of topological phases in our system. To further investigate about the topological superconducting properties of the proximity-induced AM layer, we consider the following Hamiltonian for our system,
	\begin{align}
		H(k_x,k_y) & = 
		[\xi(\textbf{k}) - \mu]\, \tau_z \sigma_0 \nonumber
		 + h_{AM}(\textbf{k})\, \tau_0 \sigma_z \nonumber \\
		& +  [ h_{rx}(\textbf{k}) \tau_z \sigma_x - h_{ry}(\textbf{k})\tau_z \sigma_y ] \nonumber \\&- \Sigma(\omega = 0)
		\label{eq.effective_ham}\ ,
	\end{align}
	where, \(\xi(\mathbf{k}) = -2t(\cos k_x + b \cos k_y)\). We introduce a dimensionless parameter \(0 < b < 1\) to represent the anisotropic hopping distortion along the \(y\)-direction. The last term $\Sigma(\omega=0)$ in Eq.~(\ref{eq.effective_ham}) represents the self-energy correction arising from the proximity-induced SC as discussed before. Interestingly, the zero-frequency ($\omega=0$) self-energy correction $\Sigma(0) = \frac{\lambda_s \Delta \tau_x}{\sqrt{\Delta^2+\eta^2}}$ appears to be independent of $\Delta$ in the limit $\eta=0$. However, this result is not counter-intuitive as ${\rm lim}_{\eta\rightarrow0} \Sigma(0)=\lambda_s \tau_x$ depends on $\Delta$ indirectly and can be understood as following. If we start with $\Delta=0$ and then consider the limit $\omega\rightarrow0$, then $\Sigma(0)$ vanishes identically. Thus, there will be no proximity-induced superconducting gap in the AM layer for $\Delta=0$ which is physically correct. Therefore, ${\rm lim}_{\eta\rightarrow0} \Sigma(0)=\lambda_s \tau_x$ is valid only for $\Delta\ne0$. In addition, the proximity-induced gap is significantly controlled by the coupling strength $\lambda_s$. In the expression, ${\rm lim}_{\eta\rightarrow0} \Sigma(0)=\lambda_s \tau_x$, $\lambda_s$ effectively mimicks the proximity-induced superconducting gap, given a nonzero value of parent gap $\Delta$~\cite{Stanesc_2017}.

	To realize the existence of topological boundary modes, we consider a ribbon geometry for our system. This is implemented by applying open boundary condition (OBC) along the finite $x$-direction, while periodic boundary condition (PBC) is employed along the $y$-direction, preserving $k_y$ as a good quantum number. The resulting energy spectrum is shown as a function of \(k_y\) in Figs.~\ref{fig:7edge_spectra_ldos}(a)–(c) for different AM-SC coupling strength $\lambda_s$. By tuning the coupling strength \(\lambda_s\), we observe a clear evolution of the energy spectrum, signaling a topological phase transition. In Fig.~\ref{fig:7edge_spectra_ldos}(a), we show the spectrum for \(\lambda_s = 0\), where the system remains gapless, consistent with the metallic nature of the bare AM layer. As \(\lambda_s\) is increased to a small finite value (\(\lambda_s = 0.2\Delta\)), a superconducting gap begins to open up at the Fermi level, as illustrated in Fig.~\ref{fig:7edge_spectra_ldos}(b). Upon further increasing the coupling to \(\lambda_s = 0.5\Delta\), the system enters a well-defined gapped TSC phase, as depicted in Fig.~\ref{fig:7edge_spectra_ldos}(c).
	Within this gap, two distinct pairs of chiral MEMs emerge, each connecting the valence and conduction bands and crossing the Fermi level at \(k_y = 0\) and \(k_y = \pm \pi\), respectively.
	The localization of these in-gap MEMs is confirmed by calculating the spatial LDOS at zero energy for a finite system (employing OBC in both directions) in the $x$-$y$ plane.
	The LDOS map, shown in Fig.~\ref{fig:7edge_spectra_ldos}(d), reveals a pronounced concentration of spectral weight at the physical boundaries of the 2D system, demonstrating that the MEMs are indeed localized at the edges. The inset of Fig.~\ref{fig:7edge_spectra_ldos}(d) further displays dispersive states at \(E_m = 0\) in the eigenvalue spectrum plotted as a function of the state index \(m\), confirming the presence of edge modes. This provides direct real-space evidence for the formation of topological boundary states.

	To characterize the topological nature of the bulk states, we compute the Chern number using the Fukui method following the prescription given in Ref.~\cite{Fukui_2005}. The Brillouin zone is discretized into a ${\bf{k}}$-space mesh, and the lattice link variables 
	are defined from the eigenstates of the occupied bands as~\cite{Fukui_2005}
	\begin{equation}
		U_\mu(\mathbf{k}) = \frac{\det \Big[ \langle u_m(\mathbf{k}) ,|, u_n(\mathbf{k}+\hat{\mu}) \rangle \Big]}
		{\left|\det \Big[ \langle u_m(\mathbf{k}) ,|, u_n(\mathbf{k}+\hat{\mu}) \rangle \Big]\right|}\ ,
		\quad \mu = x,y,
	\end{equation}
	where, ${ |u_m(\mathbf{k})\rangle }$ denotes the set of occupied eigenstates. The Berry curvature on each plaquette, $\mathbf{k}=(k_x,k_y)$, is computed using
	\begin{equation}
		F_{xy}(\mathbf{k}) = \ln \Big[ U_x(\mathbf{k}) , U_y(\mathbf{k}+\hat{x}) ,
		U_x^{-1}(\mathbf{k}+\hat{y}) , U_y^{-1}(\mathbf{k}) \Big]\ ,
	\end{equation}
	and the Chern number can be obtained by summing over the entire Brillouin zone:
	\begin{equation}
		C = \frac{1}{2\pi i} \sum_{\mathbf{k}} F_{xy}(\mathbf{k})\ .
	\end{equation}
	Our calculations reveal that the Chern number is zero throughout the $(\mu, \lambda_s)$ and $(J_a, \lambda_s)$ parameter planes. This null result is also predicted in the phenomenological case with constant $s$-wave pairing~\cite{Ghorashi_PRL}.
	
\begin{figure}
		\centering
		\includegraphics[width=1.02\linewidth]{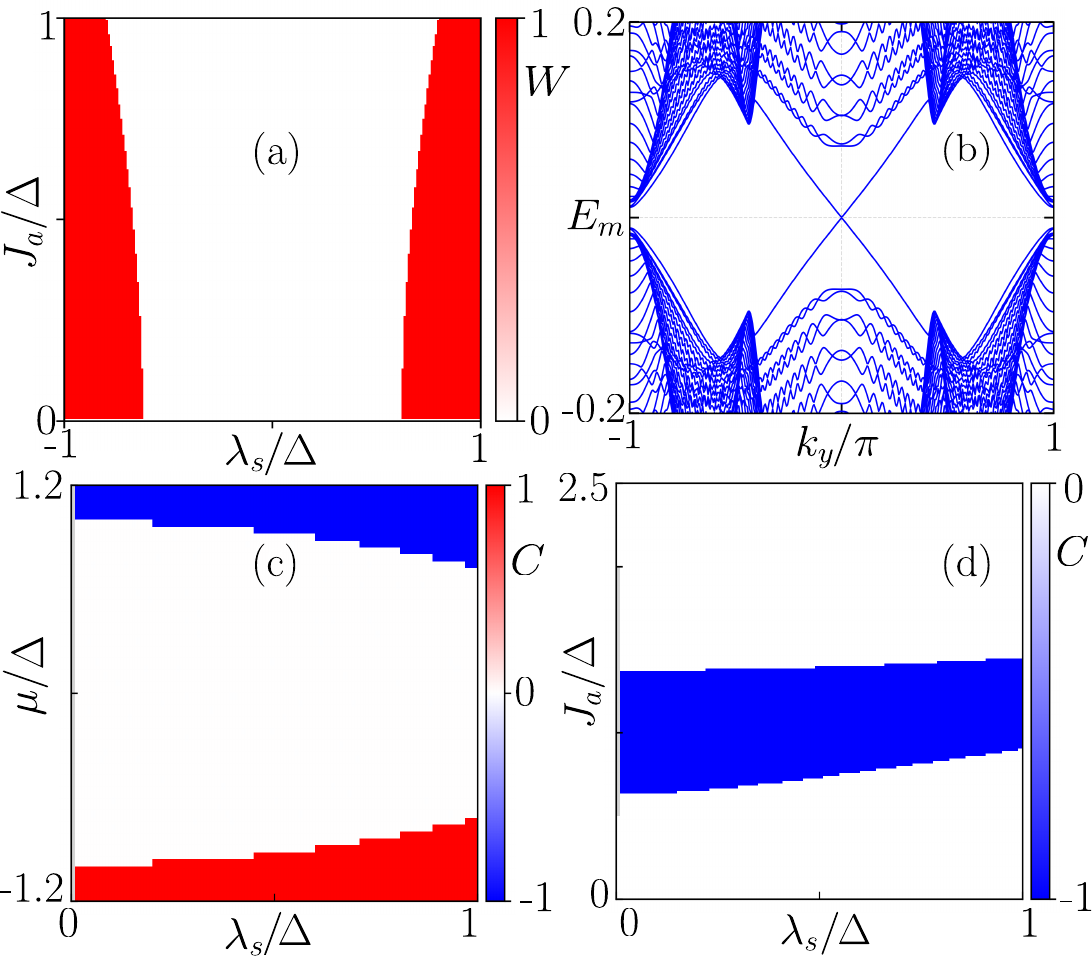} 
		\caption{Panel (a) shows the winding number $W$ in the $J_a$–$\lambda_s$ plane for the symmetric case ($b = 1$), calculated at fixed $k_y = 0$ with $\mu = 0.4t$. Panel (b) depicts the edge spectrum with OBC along the $x$-axis and PBC along the $y$-axis for the anisotropic case ($b = 0.85t$), where the gap-closing points at $k_y = \pi$ are gapped out. The chosen parameters are $\mu = 0.4t$ and $\lambda_s = 0.5\Delta$. Panel (c) displays the Chern number for the anisotropic case in the $\mu$–$\lambda_s$ plane with $J_a = 0.3t$, while panel (d) exhibits the Chern number in the $J_a$–$\lambda_s$ plane with $\mu = 0.4t$. Unless otherwise specified, the model parameters are $\Delta = 0.3t$ and $\alpha = 0.3t$.
		}
		\label{fig:topological_invariants}
	\end{figure}

		To better understand the origin of the MEMs even when the Chern number vanishes, we construct an effective 1D Hamiltonian along the high-symmetry line \(k_y = 0\). For this 1D model, which preserves chiral symmetry \(S = \tau_y \sigma_y\), the topological properties can be characterized by an integer winding number \(W\)~\cite{Ryu_2010,Chiu2016RevModPhys,Pal2025_MNLSM}. In the chiral basis where \(S\) is diagonal, the Hamiltonian takes an off-diagonal form,
		\[
		\bar{H}(k_x) = U_s^\dagger H(k_x) U_s =
		\begin{pmatrix}
			0 & q^{+}(k_x) \\
			q^{-}(k_x) & 0
		\end{pmatrix},
		\]
		where, the unitary matrix \(U_s\) is constructed from the eigenvectors of \(S\), and \(q^{\pm}(k_x)\) are the \(2\times2\) matrices corresponding to the positive and negative chiral sectors, respectively. The winding number is then defined as ~\cite{Ryu_2010,Chiu2016RevModPhys,Mondal_Rashba}
		\[
		W = \left| \pm \frac{i}{2\pi} \int_{-\pi}^{\pi} dk_x\, \mathrm{Tr}\!\left[{q^{\pm}(k_x)}^{-1}\, \partial_{k_x} q^{\pm}(k_x)\right] \right|,
		\]
		 We illustrate $W$ in Fig.~\ref{fig:topological_invariants}(a) across the \(J_a\)–\(\lambda_s\) plane, which reveals distinct regions where $W$ takes a quantized value \(W = 1\), indicating a topologically nontrivial superconducting phase hosting MEMs and protected by chiral symmetry.
	
	The coexistence of a zero Chern number ($C = 0$) with a finite winding number ($W = 1$) indicates that the system supports a weak topological superconducting (WTSC) phase~\cite{Zhu_2024}.
	Using the effective Hamiltonian in Eq.~\eqref{eq.effective_ham}, we qualitatively reinvestigate the predicted transition from a weak to a strong topological superconducting phase (STSC) with \( \lvert C \rvert = 1 \), as suggested by the phenomenological model~\cite{Ghorashi_PRL,Zhu_2024} and Eq.~\eqref{phemeno_ham} of Appendix~\ref{App.Phenomeno}. We confirm that incorporating a small anisotropic hopping term (\(b = 0.85\)) gaps out the helical edge states at \(k_y = \pm \pi\), while those at \(k_y = 0\) remain gapless dispersive, resulting in a finite Chern number of \( \lvert C \rvert = 1 \). This transition is illustrated in the edge spectra of Fig.~\ref{fig:topological_invariants}(b). Moreover, the phase diagrams depicted in Figs.~\ref{fig:topological_invariants}(c) and (d)
	demonstrate the same via the appearance of finite Chern number \( \lvert C \rvert = 1 \) in $\mu$–$\lambda_s$ and $J_a$–$\lambda_s$ planes respectively.
	
\begin{figure}
		\centering
		\includegraphics[width=1.05\linewidth]{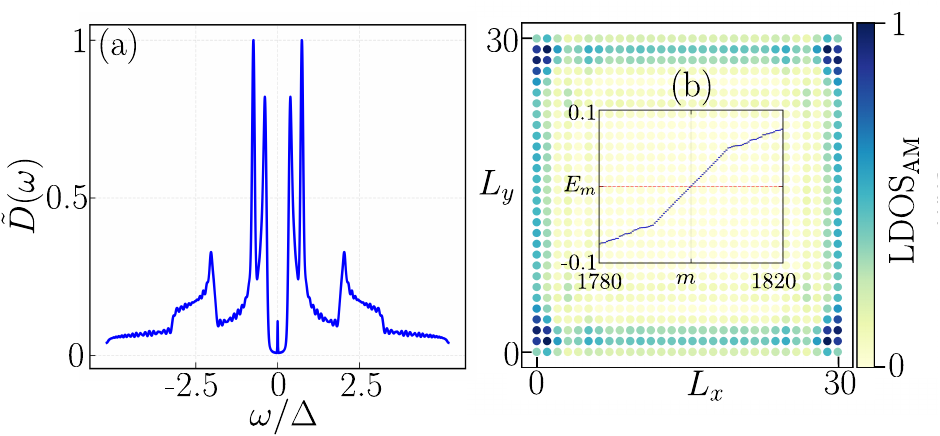}
		\caption{Panel (a) shows the energy-resolved LDOS for MEMs localized at the edges of the 2D sample. Panel (b) presents the site-resolved LDOS at zero energy for a finite size of $30 \times 30$ AM layer. The inset displays a zoomed-in view of the quasi-energy spectrum near the zero-energy. The other model parameters are chosen as $\tilde{t} = 0.5$, $\mu = 0$, $\Delta = 0.3t$, $b = 1$, $J_a = 0.1\Delta$, and $\alpha = 0.3\Delta$. These results are obtained from ED.
		}
		\label{fig:fig_9}
	\end{figure}

	
	We now compare the results obtained from our effective Hamiltonian analysis with those captured from ED as illustrated in Sec.~\ref{Sec:V_ED}. The energy-resolved LDOS for the edge modes, shown in Fig.~\ref{fig:fig_9}(a), exhibits several key features: a pronounced superconducting gap, two additional peaks-a signature of the $d$-wave AM and RSOC and a finite accumulation of spectral weight at zero energy indicating the presence of MEMs. Furthermore, the site-resolved LDOS at zero energy confirms that these MEMs are sharply localized at the physical edges of the 2D sample, as can be visualized from Fig.~\ref{fig:fig_9}(b). This spatial localization is consistent with the edge modes depicted in Fig.~\ref{fig:7edge_spectra_ldos}(d). Also, the eigenvalue spectrum highlights the gapless dispersive modes (see the inset of Fig.~\ref{fig:fig_9}(b)). 
	Therefore, the agreement between the ED results and the predictions from the effective Hamiltonian approach validates our theoretical model and the emergence of TSC phase anchoring MEMs.	
	
    \section{Summary and Conclusions}\label{Sec:conclusions}
	To summarize, in this article, we investigate superconducting proximity effects in AM-SC heterostructures using a microscopic approach. Our setup consists of a 2D altermagnetic layer placed on the surface of a three-dimensional $s$-wave SC, with direct tunneling between the first layer of the SC and the AM layer. By integrating out the superconducting degrees of freedom, we derive an effective Green's function and a corresponding low-energy effective Hamiltonian for the AM layer. Using this framework, we analyze the system's spectral features, demonstrating that a superconducting gap is indeed induced in the AM layer. The magnitude of this proximity induced gap depends on the interface coupling strength and the normal-state density of states of the SC. The resulting DOS exhibits spin-split peaks alongside characteristic superconducting coherence peaks. Furthermore, an analysis of the anomalous part of the Green's function reveals that the AM induces out-of-plane spin-triplet pairing correlations with even parity. Introducing RSOC into the AM layer generates additional in-plane triplet components with odd parity, enabling the emergence of TSC phase. 
	We further show that this effective Hamiltonian can host both WTSC and STSC phase, characterized by the presence of localized MEMs. The topological nature of these phases is established via the Chern number calculations. We also analyze the agreement of our effective Hamiltonian based results with those obtained from numerical ED. We note that in this work, we have considered two band model for the altermagnetic spin-splitting. Nevertheless, one can also consider a four band model~\cite{Smejkal_PRX_1,rasmussen2025inherentmomentumdependentgapstructure} to investigate the multi-orbital effects and  their role in enhancing the induced pairing correlations. Here, we restrict ourselves to the two band model and will investigate the orbital effect in future.
	
	
	In literature, there exist numerous studies on the superconducting proximity effect across a wide range of magnetic and non-magnetic systems. On the magnetic side, it has been extensively explored in ferromagnets (FMs)~\cite{Buzdin_2005,Halterman2001,Salamone2022,Efetov2008}, inhomogeneous FM \cite{Linder_PRB2009}, SC–FM–SC heterostructures~\cite{Golovchanskiy_2020}, and antiferromagnetic systems~\cite{Bobkova_2023,Fyhn_2023}. Although, within the present framework AM–SC and FM–SC heterostructures belong to the same symmetry classes as the magnetic order parameter is even under parity for both of them. However, the resulting proximity-induced pairing correlations exhibit qualitatively different momentum-space structures reflecting the underlying magnetic order (see Appendix.\,\ref{Sec_fm_wave} for more detailed comparision). The superconducting proximity effect has also been investigated in various non-magnetic modern platforms, such as topological insulators, Rashba nanowires, Weyl semimetals, graphene, and various mesoscopic junctions~\cite{Burset_2015, Fu_prox_2008, Khanna_2014, Bena_2013, Moriya2020, Sitthison_2014,Linder_2010, Black-Schaffer_1_2013}.
	
	Several experimental setups investigating the superconducting proximity effect have established that the interfacial coupling strength is a crucial parameter governing the induced superconducting properties inside the parent material~\cite{vanLoo2023, Flokstra2023}. In recent times, several material candidates for altermagnetism have been proposed, including \ce{RuO2}~\cite{Smejkal_PRX_1, Smejkal_PRX_2, Mazin_PRX_2022} and \ce{MnTe}~\cite{Lee2024_PRL}, as well as proximity based heterostructures \cite{zhu2025altermagneticproximityeffect}. In particular, a heterostructure combining aluminum with the recently discovered oxychalcogenide altermagnet \ce{Rb_{1-\delta}V2Te2O} exhibits a remarkably small lattice mismatch of only $0.04\%$~\cite{Ablimit_2018, Zhang2025,Heinsdorf2026}, making it a possible candidate platform for realizing the superconducting proximity effect and TSC therein studied in this work.
	

	\subsection*{Acknowledgments}
	 We acknowledge A. M. Black-Schaffer for useful comments and discussions. O.A. acknowledges K. Bera for stimulating discussions. O.A., A.P., and A.S. acknowledge SAMKHYA: High-Performance Computing Facility provided by Institute of Physics, Bhubaneswar and the two Workstations provided by Institute of Physics, Bhubaneswar from DAE APEX Project, for numerical computations. The work of PD at Physical Research Laboratory was supported by the Department of Space, Govt. of India.
	
	
	\subsection*{Data Availability Statement} The datasets generated and analyzed during the current study are available from the corresponding author upon reasonable request.

	\appendix
	\section{Derivation of the Self-Energy} \label{app:self_energy}
	Here, we compute the self-energy for an AM layer coupled via tunneling to a conventional \(s\)-wave SC, following Refs.~\cite{Khanna_2014,Bena_2013}.
	 The complete Hamiltonian is given by $H = H_{\text{AM}} + H_{\text{SC}} + H_T$, where $H_{\text{AM}}$ is given in Eq.~\eqref{eq:H_AM}, $H_{\text{SC}}$ 
	 is described by Eq.~\eqref{eq:H_SC} and $H_T$ is given in Eq.~\eqref{eq:H_T} of the main text. Here $H_{\text{SC}}$ describes a bulk $s$-wave SC. 	
	
	Using the Nambu basis, the tunneling Hamiltonian Eq.~\eqref{eq:H_T} can be compactly rewritten as
	\begin{equation} \label{eq:app_Ht_nambu}
		H_T = \sum_{r,\mathbf{R}} \Psi_{r}^\dagger \, A_{r,\mathbf{R}} \, \Phi_{\mathbf{R}} + \text{H.c.}\ ,
	\end{equation}
	where, the coupling matrix $A_{r,\mathbf{R}}$ is given by
	\begin{equation} \label{eq:app_coupling_matrix}
		A_{r,\mathbf{R}} = \ttilde \, \delta_{r, r_c} \, \delta_{\mathbf{R}, \mathbf{R}_c} \,
		\begin{pmatrix}
			1 & 0 & 0 & 0 \\
			0 & 1 & 0 & 0 \\
			0 & 0 & \text{-}1 & 0 \\
			0 & 0 & 0 & \text{-}1
		\end{pmatrix}\ .
	\end{equation}
	where $r_c=r$ is the altermagnetic layer and $\mathbf{R_c}=\mathbf{r}+\hat{z}$ is the first layer of SC.
	
	Hence, the total action for the composite system can be written as
	\begin{equation} \label{eq:app_full_action}
		S = \int_{-\infty}^{\infty} dt \left[ \sum_{\mathbf{R}} \Phi_{\mathbf{R}}^\dagger (i\hb \del_t) \Phi_{\mathbf{R}} \text{+} \sum_{r} \Psi_{r}^\dagger (i\hb \del_t) \Psi_{r} \text{-} H \right]\ ,
	\end{equation}
	where, $H$ corresponds to the total Hamiltonian. After performing a Fourier transform to frequency space ($\del_t \rightarrow -i\omega$), the action 
	can be expressed in terms of the bare Green's functions of the SC ($\Gzero_{\text{SC}}$) and the AM ($\Gzero_{\text{AM}}$) as
	\begin{align} \label{eq:app_action_freq}
		S = \int_{-\infty}^{\infty} \frac{d\omega}{2\pi} \bigg[ & \sum_{\mathbf{R,R'}} \Phi_{\mathbf{R}}^\dagger \, \left[\Gzero_{\text{SC}}(\omega)^{-1}_{\mathbf{R,R'}}\right] \Phi_{\mathbf{R'}} \nonumber \\
		+ & \sum_{r,r'} \Psi_{r}^\dagger \, \left[\Gzero_{\text{AM}}(\omega)^{-1}_{r,r'}\right] \Psi_{r'} \nonumber \\
		+ & \left( \Phi_{\mathbf{R}}^\dagger \, A_{\mathbf{R},r}^\dagger \, \Psi_{r} + \text{H.c.} \right) \bigg]\ .
	\end{align} 
	
	We now decouple the SC and AM fields by defining a new SC field $\tilde{\Phi}_{\mathbf{R}}$ through a Gaussian transformation:
	\begin{equation} \label{eq:app_field_transformation}
		\tilde{\Phi}_{\mathbf{R}} = \Phi_{\mathbf{R}} + \sum_{r, \mathbf{R'}} \Gzero_{\text{SC}}(\omega)_{\mathbf{R,R'}} \, A_{\mathbf{R'},r}^\dagger \, \Psi_{r}\ .
	\end{equation}
	Substituting this into the action~\eqref{eq:app_action_freq} yields a decoupled form given as
	\begin{align} \label{eq:app_action_decoupled}
		S = \int_{-\infty}^{\infty} \frac{d\omega}{2\pi} \bigg[ & \sum_{\mathbf{R,R'}} \tilde{\Phi}_{\mathbf{R}}^\dagger \, \left[\Gzero_{\text{SC}}(\omega)^{-1}_{\mathbf{R,R'}}\right] \tilde{\Phi}_{\mathbf{R'}} \nonumber \\
		+ & \sum_{r,r'} \Psi_{r}^\dagger \, \left[G_{\text{AM}}(\omega)^{-1}_{r,r'}\right] \Psi_{r'} \bigg]\ ,
	\end{align}
	where, $G_{AM}^{-1}(\omega)  = {G_{AM}^{0}}(\omega)^{-1} - \Sigma(\omega)$ is the inverse of the effective Green's function of the AM layer. The effect of the Cooper pair tunneling is now entirely contained within the self-energy $\Sigma$ correction terms due to SC and is given by the expression
	\begin{equation} \label{eq:app_self_energy_general}
		\Sigma_{r, r'}(\omega) = \sum_{\mathbf{R, R'}} A_{r, \mathbf{R}} \, \Gzero_{\text{SC}}(\omega)_{\mathbf{R, R'}} \, A_{\mathbf{R'}, r'}^\dagger\ .
	\end{equation}
	
	We now approximate the full SC Green's function $G_{\text{SC}}$ by its bulk value, ignoring surface effects which are not pertinent to this study. The bulk Green's function for a homogeneous BCS SC in the Nambu basis is
	\begin{equation} \label{eq:app_bulk_gf}
		\Gzero_{\text{SC}}(\omega, \mathbf{k}) = \frac{1}{\omega^2 \text{-} \xi_{\mathbf{k}}^2\text{-} \Delta^2}
		\begin{pmatrix}
			\omega \text{+} \xi_{\mathbf{k}} & 0 & \Delta & 0 \\
			0 & \omega \text{+} \xi_{\mathbf{k}} & 0 & \Delta \\
			\Delta & 0 & \omega \text{-} \xi_{\mathbf{k}} & 0 \\
			0 & \Delta & 0 & \omega \text{-}\xi_{\mathbf{k}}
		\end{pmatrix}\,
	\end{equation}
	where, $\xi_{\mathbf{k}} = \epsilon_{\text{sc}} - 2t_{\text{sc}} \sum_{i=x,y,z} \cos(k_i a)$ is the normal-state dispersion relation ($a$ is the lattice constant). In real space, this can be written as
	\begin{equation}
		\Gzero_{\text{SC}}(\omega)_{\mathbf{R,R'}} = \frac{1}{N} \sum_{\mathbf{k}} \Gzero_{\text{SC}}(\omega, \mathbf{k}) \, e^{i\mathbf{k} \cdot (\mathbf{R} - \mathbf{R'})}\ .
	\end{equation}
	
	Substituting Eqs.~\eqref{eq:app_coupling_matrix} and~\eqref{eq:app_bulk_gf} into the general self-energy formula~\eqref{eq:app_self_energy_general}, and noting that the coupling is local (\ie $r=r'=r_c$, $\mathbf{R}=\mathbf{R'}=\mathbf{R}_c$), we find the on-site self-energy as
	\begin{align} \label{eq:app_sigma_int}
		&\Sigma_{r_c}(\omega) = |\ttilde|^2 \, \Gzero_{\text{SC}}(\omega)_{\mathbf{R}_c, \mathbf{R}_c} \nonumber \\
		& = |\ttilde|^2 \frac{1}{N} \sum_{\mathbf{k}} \frac{ \omega \tauzero \szero + \Delta \tax \szero + \xi_{\mathbf{k}} \tauz \szero }{\omega^2 - \xi_{\mathbf{k}}^2 - \Delta^2}\ .
	\end{align}
	
    Converting the sum over $\mathbf{k}$ to an integral over energy $\xi$ weighted by the normal density of states at the Fermi level $N(0)$, and ignoring the terms odd in $\xi$,  we obtain the final expression for the self-energy as
	\begin{align} \label{eq:app_sigma_final}
		\Sigma(\omega) &= \delta_{r, r_c} \, \pi N(0) |\ttilde|^2 \int_{-\infty}^{\infty} d\xi \, \frac{ \omega \tauzero \szero + \Delta \tax \szero }{\omega^2 - \xi^2 - \Delta^2} \nonumber \\
		&= \, \frac{\pi N(0) |\ttilde|^2}{\sqrt{\Delta^2 - \omega^2}} \left( \omega \tauzero \szero + \Delta \tax \szero \right)\ .
	\end{align}
	with $\lambda_{s}=\pi N(0) |\ttilde|^2$. This is the result used in the main text. The branch cut of the square root in the denominator is chosen such that $\Im[\Sigma] < 0$ for retarded functions.
	
		
	\section{Limit of Exchange Energy in Spectral Analysis (DOS)}\label{Appendix_limit}
	We have investigated the Green's functions and DOS at $\omega=0$ using Eq.~\eqref{eq:Green_function} and \eqref{eq:DOS} of the main text. This allows one
	to understand the critical behaviour mentioned in Sec.~\ref{Sec:III_dos}, as finite values of DOS at $\omega=0$ (since $\mu=0$) suggests a gapless phase otherwise gapped phase. Analytically, the integral of DOS takes the form at $\omega=0$ as,
	\[
	D(\eta,\omega=0) = -\frac{1}{\pi}\iint_{\text{BZ}} \frac{\eta \, f(k_x, k_y)}{g(k_x, k_y) + h(k_x, k_y) \, \eta^2} \frac{ \, dk_x \, dk_y}{2\pi^2}
	\]
	in the limit $\eta \to 0^+$, where
	\begin{align*}
		f &= h_{AM}(\textbf{k})^2 + h_t(\textbf{k})^2 B^2 + \lambda_s^2\ , \\
		g &= h_{AM}(\textbf{k})^2 - h_t(\textbf{k})^2 B^2 - \lambda_s^2\ , \\
		h &= 4 h_{AM}(\textbf{k})^2\ ,
	\end{align*}
	Here, $h_{AM}(\textbf{k})$, $h_t(\textbf{k})$ and $\lambda_s$ are defined in the main text with $J_{a}, t, \lambda_s > 0$.
	
	The behavior of $D(\eta)$ as $\eta \to 0$ changes dramatically at a critical value $J_{a}\equiv j_{c}=\lambda_s/2$: for $J_{a} < j_c$, $\lim_{\eta \to 0} D(\eta) = 0$, while for $J_{a} > j_c$, $\lim_{\eta \to 0} D(\eta)$ becomes finite and nonzero.
	The denominator of the integrand $g + h \eta^2$ can become small of order $\eta^2$ near curves where $g(k_x, k_y) = 0$, yielding a finite contribution as $\eta \to 0$. This occurs only if $g(k_x, k_y) = 0$ has solutions in the Brillouin zone. Choosing $(U,V)$ such that $U = (\cos k_x + \cos k_y)/\sqrt{2}$, 
	$V = (\cos k_x - \cos k_y)/\sqrt{2}$, we have a solution
	\[
	g = 2J_{a}^{2} V^2 - 8t^2 U^2 - \lambda_s^2.
	\]
	The maximum occurs at $V = \pm \sqrt{2}$, $U = 0$ (\ie $(\cos k_x, \cos k_y) = (1,-1)$ or $(-1,1)$), yielding $g_{\text{max}} = 4J_{a}^2 - \lambda_s^2$. 
	Thus if $4J_{a}^2 - \lambda_s^2 < 0$ ($J_{a} < \lambda_s/2$), then $g(k_x, k_y) < 0$ throughout the Brillouin zone, while if $4J_{a}^{2} - \lambda_s^2 > 0$ 
	($J_{a} > \lambda_s/2$), then $g(k_x, k_y) > 0$ in some region in the zone.
	
	When $g < 0$ everywhere, the denominator of the integrand remains $O(1)$ as $\eta \to 0$, making the integrand $O(\eta)$ and the integral vanishes. When $g$ changes sign, the curve $g = 0$ exists, and near this curve the denominator $\sim h \eta^2$, making the integrand $O(1/\eta)$ in a region of width $O(\eta)$, giving rise to a finite contribution. The critical value $j_c = \lambda_s/2$ therefore separates the gapped ($J_{a} < j_c$) and gapless ($J_{a} > j_c$) phases.

	\section{Phenomenological Hamiltonian and Green's Function}\label{App.Phenomeno}
	The phenomenological Hamiltonian corresponding to momentum space can be written in the BdG basis as~\cite{Ghorashi_PRL}, 
	\begin{widetext}
		
		\begin{equation}
			H(k_x, k_y) = -2t(\cos k_x + b \cos k_y) \tau_0 \sigma_0 - \mu \tau_z \sigma_0 + J_a (\cos k_x - \cos k_y) \tau_0 \sigma_z + 2\alpha (\sin k_y \tau_z \sigma_x - \sin k_x \tau_z \sigma_y) + \Delta \tau_x \sigma_0\ ,
			\label{phemeno_ham}
		\end{equation}		
	\end{widetext}
	where, the parameters, $\mu$, $t$, $\alpha$, $J_a$, and $\Delta$ carry their usual meaning as defined in the main text. The DOS can be calculated using the imaginary part of phenomenological retarded Green's function, given by:
	\begin{equation}
		G^{-1}(\omega, \mathbf{k}) = (\omega + i \eta) I - H(\mathbf{k})\ ,
	\end{equation}
	Here, $H(\mathbf{k})$ is defined in Eq. \eqref{phemeno_ham}, $\eta$ is a small positive parameter, and $I$ is the identity matrix. We investigate the spectral properties presented in the main text without the RSOC ($\alpha=0$).
	
	\begin{figure*}[htbp]
		\centering
		\includegraphics[width=\textwidth]{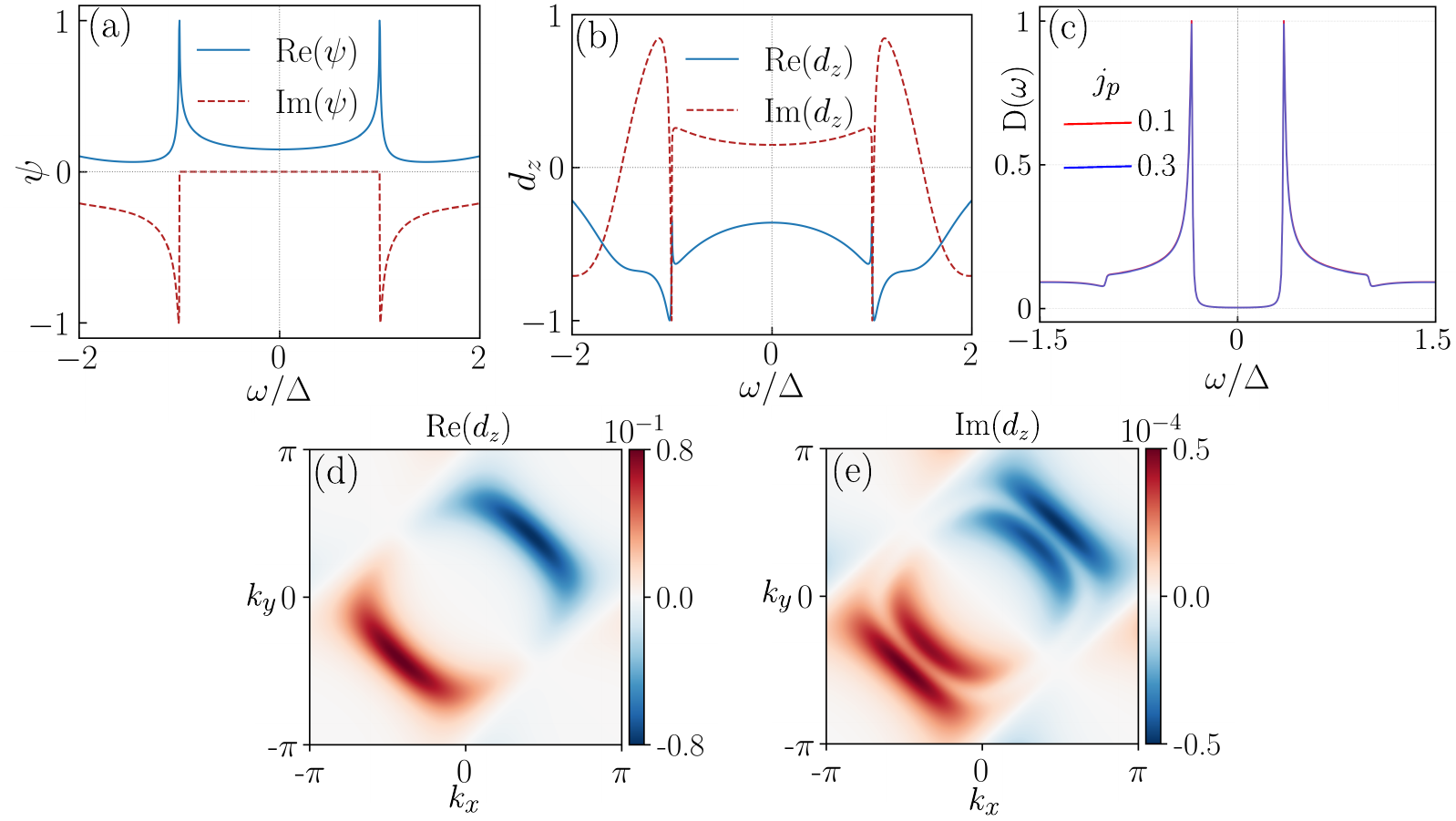}
		\caption{
			(a) Singlet component $\psi(\omega)$ is shown as a function of frequency for an arbitrary fixed momentum.
			(b) Triplet component $d_z(\omega)$ is depicted as a function of frequency choosing an arbitrary fixed momentum.
			(c) Density of states (DOS) is displayed for two values of $p$-wave exchange coupling strength $J_p$.
			(d) Real part of the triplet component $\text{Re}[d_z(\mathbf{k},\omega)]$ is shown in the $k_x$-$k_y$ plane at $\omega = 0.5\Delta$.
			(e) Imaginary part of the triplet component $\text{Im}[d_z(\mathbf{k},\omega)]$ is
			depicted in the $k_x$-$k_y$ plane choosing $\omega = 0.5\Delta$.
			Other model parameters are considered as $\Delta = 0.3t$, $\mu = 0$, $\lambda_s = 0.4\Delta$.
		}
		\label{fig:fig_10}
	\end{figure*}
	\section{Analysis of the pairing symmetry for $p$-wave magnets}\label{Sec_p_wave}
	Here, we explore the symmetry of the pair amplitudes proximitized to a $p$-wave magnet to elucidate the odd-parity nature of its triplet components. The model Hamiltonian for the $p$-wave magnet is given by~\cite{Linder_p_magnet_PRL,Nagae_2025,Hellenes2024pwavemagnets,Maeda_2024_p_magnet},
	\begin{equation}
		H_\text{PM} = h_t(\textbf{k}) \tau_z \sigma_0
		+ J_p \left( \sin k_x + \sin k_y \right) \tau_z \sigma_z\ ,
		\label{eq:H_PAM}
	\end{equation}
	where, $h_t(\textbf{k})$ is defined in the main text and $J_p$ denotes the exchange coupling of $p$-wave magnet.
	The effective Green's function for the $p$-wave magnetic layer can be written in a form analogous to Eq.~\eqref{eq:Green_function} as
	\begin{equation*}
		G_{PM}(\omega, k_x, k_y) = \left[ (\omega + i\eta)\, I - H_{\text{PM}}(k_x, k_y) - \Sigma(\omega) \right]^{-1}\ ,
		\label{eq:G_pam}
	\end{equation*}
    This is a $4 \times 4$ matrix in Nambu space. Following Eq.~\eqref{eq:G_matrix} of the main text, the anomalous parts of the Green's function are off-diagonal $2 \times 2$ matrices in spin space and can be written as,
   \begin{equation*}
   \bar{\Delta} = \begin{pmatrix}
   	\dfrac{\text{-}q(\omega)}{\Lambda(\mathbf{k},\omega) \text{-} 2 h_{p}(\textbf{k}) h_{t}(\textbf{k})} & 0 \\
   	0 & \dfrac{\text{-}q(\omega)}{\Lambda(\mathbf{k},\omega) \text{+} 2 h_{p}(\textbf{k}) h_{t}(\textbf{k})}
   \end{pmatrix}\ ,
   \end{equation*}
   where, the auxiliary term is defined as $\Lambda(\textbf{k},\omega)=h_p(\textbf{k})^2+h_t(\textbf{k})^2-p(\omega)^2+q(\omega)^2$. Here, $h_p(\textbf{k})=J_p ( \sin k_x + \sin k_y )$ originates from the $p$-wave magnet, while $h_t(\textbf{k})$, $p(\omega)$, and $q(\omega)$ correspond to as defined in the main text. This $2 \times 2$ matrix can now be decomposed according to Eq.~\eqref{eq:decouple} to yield the triplet and singlet components as follows:
	\begin{equation}
	\begin{aligned}
		\psi(\omega, \mathbf{k}) &= 
		\frac{\text{-}q(\omega) \Lambda(\omega, \mathbf{k})}{\Lambda(\omega, \mathbf{k})^2 - 4 h_{p}(\mathbf{k})^2 h_t(\textbf{k})^2}\ , \\
		d_z(\omega, \mathbf{k}) &= 
		\frac{\text{-}2 h_{p}(\mathbf{k}) h_t(\textbf{k}) q(\omega)}{\Lambda(\omega, \mathbf{k})^2 - 4 h_{p}(\mathbf{k})^2 p(\omega)^2}\ , \\
		& d_x(\omega, \mathbf{k})  = d_y(\omega, \mathbf{k}) = 0\ .	 
	\end{aligned}
	\label{eq:pair_components_pm}
\end{equation}

	The analytical expressions derived above (Eq.~\eqref{eq:pair_components_pm}) manifest that only the singlet $\psi$ and the $z$-polarized triplet $d_z$ components are finite for $p$-wave magnet, whereas $d_x$ and $d_y$ vanish identically. The denominator $\Lambda(\omega, \mathbf{k})^2 - 4 h_{p}(\mathbf{k})^2 p(\omega)^2$ is an even function of momentum since $\Lambda^2$ and $h_p^2$ are even in $\textbf{k}$. Hence, the parity of each component is governed entirely by the numerator. For the singlet component $\psi$, the numerator contains only even functions of momentum, making it even under inversion $\mathbf{k}\to -\mathbf{k}$. In contrast, the numerator of the triplet component $d_z$ is proportional to $h_p(\mathbf{k})h_t(\mathbf{k})$, where $h_p(\mathbf{k})$ is odd in momentum, leading to an overall odd-parity triplet state. These features are illustrated in Fig.~\ref{fig:fig_10}. In frequency space, both components are even functions of $\omega$: the singlet $\psi(\omega)$ and the triplet $d_z(\omega)$ exhibit even-frequency behavior (see Figs.~\ref{fig:fig_10}(a),(b)). 
	On the other hand, singlet $\psi(\omega)$ and the triplet $d_z(\omega)$ components reveal even and odd parity feature respectively as depicted in Figs.~\ref{fig:fig_10}(d),(e). 
	Thus, the singlet belongs to the ESE class, while the triplet $d_z$ belongs to the ETO class. 
	
	The DOS of a proximity induced $p$-wave magnetic layer is shown in 
	Fig.~\ref{fig:fig_10}(c). The latter manifests that, for different values of the $p$-wave exchange order $J_p$, no superconducting gap suppression or Meissner effect-like features are observed. This indicates that for the odd-parity pairing realized here, TRS remains preserved. Consequently, within the present model the system remains topologically trivial, despite the presence of odd-parity pairing. To realize a nontrivial topological phase in this system, an additional time-reversal symmetry breaking element such as external Zeeman field is required that explicitly lifts the spin degeneracy as shown in  Ref.\,\cite{sun2025}. We believe that in the proximity scenario as well, nontrivial topological phase can be realized in $p$-wave magnets with an in-plane Zeeman field.

\section{Comparison with the pairing symmetry for ferromagnet}\label{Sec_fm_wave}
\begin{figure}
	\centering
	\includegraphics[width=0.98\columnwidth]{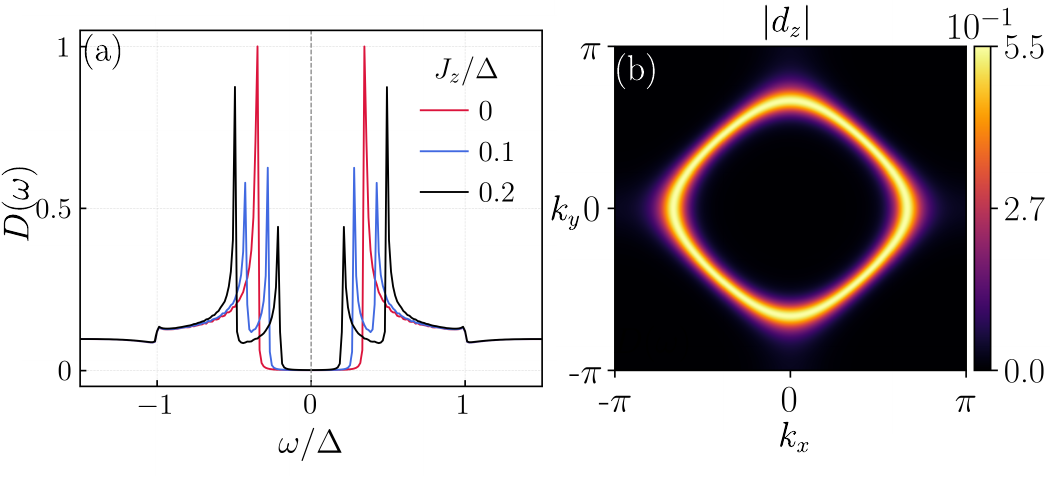}
	\caption{In panel (a), we present the DOS $D(\omega)$ of the FM layer at a fixed coupling strength $\lambda_s = 0.5\Delta$ for different values of $J_z$. Panel (b) depicts the absolute value of the triplet component $d_z$ in the $k_x$-$k_y$ plane at $\omega = 0.5\Delta$. The other model parameters are chosen as $\mu = 0$, $\Delta = 0.3t$, $\lambda_s = 0.5\Delta$, and $\alpha = 0$.
	}
	\label{fig:DOS_FM}
\end{figure}
In this appendix, we explore the symmetry of the pair amplitudes proximatized to the FM in the presence and absence of an interfacial RSOC for a comparison with our results for the AM. The model Hamiltonian we consider can be written as, 
\begin{align}
	H_{\mathrm{FM+RSOC}}(\mathbf{k}) 
	& = h_t(\textbf{k})\,\tau_z\sigma_0 
	+J_z\,\tau_0\sigma_z \nonumber\\ 
	&+ 2\alpha(\sin k_y\,\tau_z\sigma_x - \sin k_x\,\tau_z\sigma_y)\ ,
	\label{eq:H_FM_RSOC}
\end{align}
where, all the terms carry the same meaning as mentioned in Eq.~\eqref{eq:H_AM_RSOC} in the main text, and $J_z$
denotes the magnetic exchange field arising from the FM.
The effective Green's function for the ferromagnetic layer in presence of interfacial RSOC can be written in a form analogous to Eq.~\eqref{eq:Green_function} as
\begin{equation}
	G_{FM}(\omega, k_x, k_y) = \left[ (\omega + i\eta)\, I - H_{\text{FM+RSOC}}(k_x, k_y) - \Sigma(\omega) \right]^{-1}.	\label{eq:G_fm-rsoc}
\end{equation}
We use this effective Green’s function to analyze the total DOS of the SM. Furthermore, from the anomalous part of the Green’s function, as prescribed in Sec.~\ref{Sec:III_dos}, we compute the proximity-induced singlet and triplet pair amplitudes for the ferromagnetic layer in the presence and absence of an interfacial RSOC, and analyze their symmetry properties.

First, we consider the case when there is no interfacial RSOC term, i.e., $\alpha = 0$.
The variation of the DOS as a function of energy is presented in Fig.~\ref{fig:DOS_FM}(a). It illustrates the evolution of the DOS as one varies the FM exchange coupling $J_z$, at a fixed coupling strength $\lambda_s = 0.5\Delta$, where the superconducting gap is already induced. As $J_z$ increases, the proximity-induced gap is gradually suppressed, signalling the competition between superconductivity and the FM exchange field. For finite values of $J_z$, the coherence peaks split, reflecting the spin splitting of the quasiparticle spectrum. This behavior is qualitatively analogous to the AM case [see Fig.~\ref{fig:fig_2}]. However, an important quantitative difference arises as in case of the AM, the maximum exchange energy originating from the term $J_a(\cos k_x - \cos k_y)$ is $2J_a$, whereas in the FM case the exchange splitting is simply $J_z$. Consequently, the coherence peak splitting in the AM case becomes twice that of the FM case when comparing for equal values of the corresponding coupling strengths.

Then, the singlet and triplet components are obtained analytically, from the anomalous part of the effective Green's functions (Eq.~\eqref{eq:G_fm-rsoc}) for the case without interfacial RSOC ($\alpha$=0) as,
\begin{equation}
	\begin{aligned}
		\psi(\omega, \mathbf{k}) &= 
		\frac{q(\omega)\,\tilde{\Xi}(\omega, \mathbf{k})}{\tilde{\Xi}(\omega, \mathbf{k})^2 - 4 J_z^2 p(\omega)^2}\ , \\
		d_z(\omega, \mathbf{k}) &= 
		\frac{2 J_z\, p(\omega)\, q(\omega)}{\tilde{\Xi}(\omega, \mathbf{k})^2 - 4 J_z^2 p(\omega)^2}\ ,
		d_x(\omega, \mathbf{k}) = d_y(\omega, \mathbf{k}) = 0\ ,
	\end{aligned}
	\label{eq:pair_components_FM}
\end{equation}
where, the auxiliary term $\tilde{\Xi}(\omega, \mathbf{k})$ is defined as, $\tilde{\Xi}(\omega, \mathbf{k})=J_z^2-h_t^2+p^2-q^2$, all the other components are already defined in the main text. The analytical expressions derived above (Eq.~\eqref{eq:pair_components_FM}) manifest that only the singlet $\psi$ and the $z$-polarized triplet $d_z$ components are finite for FM, whereas $d_x$ and $d_y$ vanish identically. The denominator $\tilde{\Xi}(\omega, \mathbf{k})^2 - 4 J_z^2 p(\omega)^2$ is an even function of momentum since $\tilde{\Xi}(\omega, \mathbf{k})$ is even in $\textbf{k}$. Hence, the parity of each component is governed entirely by the numerator. For the singlet component $\psi$, the numerator contains only even functions of momentum, making it even under inversion $\mathbf{k}\to -\mathbf{k}$. In contrast, the numerator of the triplet component $d_z$ is proportional to $J_z$, which is independent of momentum, leading to an overall even-parity triplet state. For the frequency dependence, since the expressions are exactly the same as in the AM case [Eq.~\eqref{eq:pair_components}], without any ambiguity we can conclude that the singlet component is even in frequency, while the triplet component is odd in frequency. Thus, the singlet belongs to the ESE class, while the triplet $d_z$ belongs to the OTE class. Importantly, here the $d_z(\mathbf{k}, \omega)$ component originates from the FM exchange coupling, as it depends directly on $J_z$ in its analytical form. Moreover, if we inspect the absolute value of the frequency-resolved OTE component $|d_z|$ in Fig.~\ref{fig:DOS_FM}(b), we observe that it does not exhibit any nodal structure, in contrast to the AM case.

Now, if we turn on the interfacial RSOC i.e., $\alpha \neq 0$, we obtain the analytical expressions for the singlet and triplet components of the pair amplitudes as,
\begin{figure*}[t]
	\centering
	\includegraphics[width=1.02\textwidth]{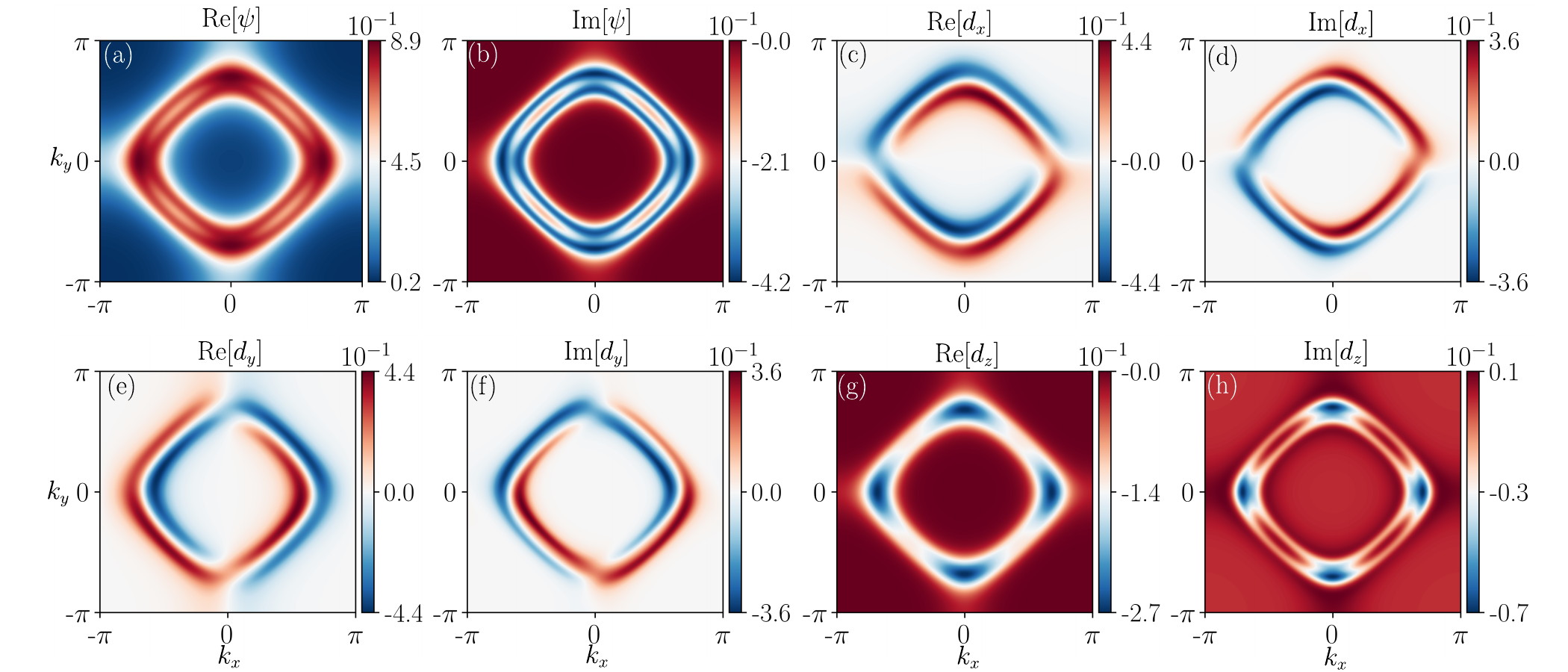}
	\caption{Momentum-resolved pairing amplitudes are depicted in the $k_{x}$-$k_{y}$ plane choosing $\omega = 0.5\Delta$. In panels (a) and (b) we show the 
		real and imaginary parts of the singlet component $\psi(\mathbf{k})$. On the other hand, panels (c)–(h) display the real and imaginary parts of the triplet components $d_x(\mathbf{k})$, $d_y(\mathbf{k})$, and $d_z(\mathbf{k})$. The finite in-plane triplet components $d_x$, $d_y$ arise in the presence of RSOC [$\alpha=0.3\Delta$]. The other model parameters used are $J_z=0.1\Delta$, $\mu=0$, $\Delta=0.3 t$, $\lambda_s=0.5 \Delta$.}
	\label{fig:fig_11}
\end{figure*}
\begin{equation}
	\begin{aligned}
		\psi(\omega, \mathbf{k}) &= 
		\frac{q \left( h_{\mathrm{rx}}^2 + h_{\mathrm{ry}}^2 + h_t^2 - J_z^2 - p^2 + q^2 \right)}
		{\Upsilon(\mathbf{k},\omega)}\ , \\
		d_x(\omega, \mathbf{k}) &= 
		\frac{2q \left( i h_{\mathrm{rx}} J_z - h_{\mathrm{ry}} h_t \right)}
		{\Upsilon(\mathbf{k},\omega)}\ , \\
		d_y(\omega, \mathbf{k}) &= 
		\frac{2q \left( h_{\mathrm{rx}} h_t + i h_{\mathrm{ry}} J_z \right)}
		{\Upsilon(\mathbf{k},\omega)}\ , \\
		d_z(\omega, \mathbf{k}) &= 
		-\frac{2p q J_z}{\Upsilon(\mathbf{k},\omega)}\ .
	\end{aligned}
	\label{eq:pair_components_rsoc_1}
\end{equation}
where, the denominator is given by
\begin{widetext}
	\begin{multline}
		\Upsilon = 
		J_z^4 + h_{\mathrm{rx}}^4 + h_{\mathrm{ry}}^4 
		- 2 h_{\mathrm{ry}}^2 h_t^2 + h_t^4
		- 2 h_{\mathrm{ry}}^2 p^2 - 2 h_t^2 p^2 + p^4
		+ 2 h_{\mathrm{ry}}^2 q^2 + 2 h_t^2 q^2 - 2 p^2 q^2 + q^4 \\
		+ 2 J_z^2 ( h_{\mathrm{rx}}^2 + h_{\mathrm{ry}}^2 - h_t^2 - p^2 - q^2 )
		+ 2 h_{\mathrm{rx}}^2 ( h_{\mathrm{ry}}^2 - h_t^2 - p^2 + q^2 )\ ,
		\label{eq:Gamma}
	\end{multline}
\end{widetext}
where, $h_{\mathrm{rx}}(\mathbf{k})$, $h_{\mathrm{ry}}(\mathbf{k})$, $p(\omega)$, $q(\omega)$, and $h_t(\mathbf{k})$ are defined in Sec.~\ref{Sec:IV_Symm_RSOC} and in Eq.~\eqref{eq:Auxiliary_terms}.
\begin{table}
	\centering
	\small
	\renewcommand{\arraystretch}{1.2}
	\caption{Classification of singlet and triplet pairing symmetries 
		in $p$-wave magnet/$s$-wave SC and FM /$s$-wave SC hybrid system with and without RSOC.}
	\vspace{0.2cm}
	\begin{tabular}{|c|c|c|c|c|}
		\hline
		System & Pairing & Frequency & Parity & Symmetry \\
		\hline
		\multirow{2}{*}{\makecell{$p$-wave magnet \\ + $s$-wave SC}} 
		& Singlet ($\psi$) & Even & Even & ESE \\
		\cline{2-5}
		& Triplet ($d_z$) & Even & Odd & ETO \\
		\hline
		\multirow{2}{*}{\makecell{FM \\ + $s$-wave SC}} 
		& Singlet ($\psi$) & Even & Even & ESE \\
		\cline{2-5}
		& Triplet ($d_z$) & Odd & Even & OTE \\
		\hline
		\multirow{3}{*}{\makecell{FM \\ + $s$-wave SC \\ + RSOC}} 
		& Singlet ($\psi$) & Even & Even & ESE \\
		\cline{2-5}
		& Triplet ($d_z$) & Odd & Even & OTE \\
		\cline{2-5}
		& Triplet ($d_x,d_y$) & Even & Odd & ETO \\
		\hline
	\end{tabular}\label{Table:II}
\end{table}
Since the denominator $\Upsilon(\mathbf{k},\omega)$ is an even function of momentum, the parity of each pairing component is determined solely by its numerator. Both the singlet $\psi$ and the out-of-plane triplet $d_z$ components are even in momentum, while the in-plane triplet components $d_x$ and $d_y$ are odd in momentum due to the RSOC terms $h_{\mathrm{rx}}$ and $h_{\mathrm{ry}}$, respectively. 
This momentum dependence is clearly visible in Fig.~\ref{fig:fig_11}, which displays the real and imaginary parts of the frequency-resolved singlet and triplet components in the $k_x$–$k_y$ plane. From Figs.~\ref{fig:fig_11}(a), (b), (g), and (h), it is evident that the singlet $\psi$ and the triplet component $d_z$ exhibit even parity in momentum space. In contrast, Figs.~\ref{fig:fig_11}(c)–(f) exhibit that the triplet components $d_x$ and $d_y$ display odd parity. Notably, these odd-parity triplet components are absent without RSOC but appear once an interfacial  RSOC is introduced.
Although the parity of the singlet and triplet components is exactly the same as in the AM case in presence of interfacial RSOC, the nodal lines that appear in the AM system (see Fig.~\ref{fig:fig_4})-visible in the real parts of the $d_x$ and $d_y$ triplet components as well as in $d_z$-are absent in the FM case with interfacial RSOC.
In this case, for the frequency dependence, since the expressions of the pair amplitudes are exactly the same as in the AM case in the presence of RSOC [Eq.~\eqref{eq:pair_components_rsoc}], except that $h_{AM}(\mathbf{k})$ is replaced by $J_z$, we can conclude without ambiguity that the singlet component is even in frequency, while the triplet component $d_z$ is odd in frequency, and the triplet components $d_x$ and $d_y$ are even in frequency, identical to the AM case. Thus, the singlet belongs to the ESE class, the triplet $d_z$ component belongs to the OTE class, and the in-plane triplet components ($d_x$ and $d_y$) belong to the ETO class.

In conclusion, the FM system exhibits the same pairing symmetry classes as the AM system, both in the presence and absence of interfacial RSOC, with the only differences arising from their distinct momentum-space structures.  In the Table~\ref{Table:II}, we extend the symmetry classification of the induced pair amplitudes for both the $p$-wave magnet and the FM cases.

\begin{figure}
	\centering
	\includegraphics[width=1.05\columnwidth]{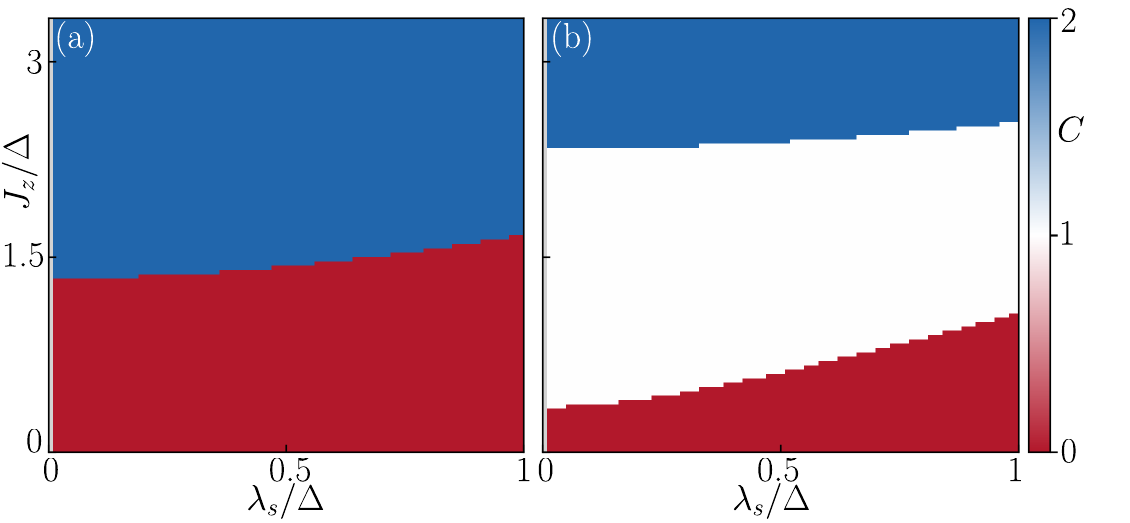} 
	\caption{ In panel (a), we display the the Chern number ($C$) for the isotropic case ($b=1$) for the proximatized FM layer with interfacial RSOC, in the $J_z$–$\lambda_s$ plane while panel (b) exhibits the Chern number ($C$) for the anisotropic case ($b=0.85$) in the $J_z$–$\lambda_s$ plane with $\mu = -0.4t$. The other model parameters are considered as $\Delta = 0.3t$ and $\alpha = 0.3t$.
	}
	\label{fig:Fig13}
\end{figure}
To further investigate the topological superconducting properties of the proximity-induced FM layer with RSOC, we consider the following effective model Hamiltonian similar to Eq.~\eqref{eq.effective_ham} as,
\begin{align}
	H_{\mathrm{FM+RSOC}}(\mathbf{k}) 
	& = [\xi_{\mathbf{k}}-\mu]\,\tau_z\sigma_0 
	+J_z\,\tau_0\sigma_z \nonumber\\ 
	&+ 2\alpha(\sin k_y\,\tau_z\sigma_x - \sin k_x\,\tau_z\sigma_y)\\
	&+\Sigma(\omega=0) \nonumber
	\label{eq:H_FM_RSOC_topo}
\end{align}
where, all the components are defined in Eq.~\eqref{eq.effective_ham}. To characterize the topological nature of the underlying bulk states, we compute the Chern number using the Fukui method following the prescription given in the main text and in Ref.~\cite{Fukui_2005} for two different cases. They are when the hopping term is isotropic, \ie $b=1$ and when there is a small hopping anisotropy present, \ie $b=0.85$. In Fig.~\ref{fig:Fig13}(a), we present the Chern number $C$ in the $J_z-\lambda_s$ plane for the isotropic hopping case $b=1$. We identify a topological superconducting phase (STSc) with $C=2$ (blue region) and a trivial phase with $C=0$ (red region). This behavior is qualitatively different from the AM system as in the AM case, isotropic hopping does not support a $C=2$ topological superconducting phase; instead, it hosts a WTSc phase. In Fig.~\ref{fig:Fig13}(b), we showcase the Chern number $C$ in the $J_z\!-\!\lambda_s$ plane for the anisotropic hopping case $b=0.85$. In this case, apart from the trivial phase $C=0$ (red region), we identify two distinct topological phases with $C=1$ (white region) and $C=2$ (blue region) which are STSc. This behaviour is also qualitatively different from the AM system: in the AM case, anisotropy in hopping does not give rise to a $C=2$ STSc phase; instead, it hosts a STSc phase with $C=1$. Therefore, proximity-induced FM layer with RSOC can give rise to distinct topological phases compared to the AM layer with interfacial RSOC and proximity-induced superconductivity.
	\bibliography{references.bib}
\end{document}